\documentclass[twocolumn,showpacs,aps,pra,eqsecnum]{revtex4}
\usepackage{graphicx}
\usepackage{dcolumn}
\usepackage{bm}
\usepackage{amsfonts}
\usepackage{latexsym}
\usepackage{amssymb} 
\usepackage{verbatim}
\usepackage{amsthm}
\usepackage{amsmath}

\def\fun#1#2{\lower3.6pt\vbox{\baselineskip0pt\lineskip.9pt
  \ialign{$\mathsurround=0pt#1\hfil##\hfil$\crcr#2\crcr\sim\crcr}}}
\newcommand{\NN}{{{\mathbb N}}}

\newcommand{\TR}{{{\mbox{Tr}}}}
\begin{document}


\title{
  All-order evaluation of weak measurements\\
  {\small --- The cases of an operator ${\bf A}$ which satisfies 
    the property ${\bf A}^{2}=1$ ---}
}

\author{
  Kouji Nakamura${}^{1}$\footnote{E-mail address: kouji.nakamura@nao.ac.jp},
  Atsushi Nishizawa${}^{2}$\footnote{E-mail address: anishi@yukawa.kyoto-u.ac.jp},
  and
  Masa-Katsu Fujimoto${}^{1}$\footnote{E-mail address: fujimoto.masa-katsu@nao.ac.jp}
}
\affiliation{%
  ${}^{1}$TAMA project, Optical and Infrared Astronomy Division,  
  National Astronomical Observatory of Japan,
  Mitaka, Tokyo 181-8588, Japan\\
  ${}^{2}$Yukawa Institute for Theoretical Physics, Kyoto University,
  Kyoto 606-8502, Japan   
}%


\date{\today}

\begin{abstract}
  Some exact formulae of the expectation values and
  probability densities in a weak measurement for an operator 
  ${\bf A}$ which satisfies the property ${\bf A}^{2}=1$ are
  derived. 
  These formulae include all-order effects of the unitary
  evolution due to the von-Neumann interaction.
  These are valid not only in the weak measurement regime but
  also in the strong measurement regime and tell us the
  connection between these two regime.
  Using these formulae, arguments of the optimization of the
  signal amplification and the signal to noise ratio are
  developed in two typical experimental setups.
\end{abstract}

\maketitle

\section{Introduction}
\label{sec:intro}


Since the proposal of the weak measurement by Aharonov, Albert,
and Vaidman (AAV)~\cite{Y.Aharonov-D.Z.Albert-L.Vaidman-1988} in
1988, weak measurements have been investigated by many
researchers. 
The idea of weak measurement has been used to resolve
fundamental paradoxes in quantum mechanics such as Hardy's 
paradox~\cite{Y.Aharonov-A.Botero-S.Pospescu-B.Reznik-J.Tollaksen-2001}.
In addition to many theoretical works on weak measurements, 
it is important to note that some experiments realized this weak
measurement in different experimental
setups~\cite{N.W.M.Ritchie-J.G.Story-R.G.Hulet-1991,G.J.Pryde-J.L.O'Brien-A.G.White-T.C.Ralph-H.M.Wiseman-2005,O.Hosten-P.Kwiat-2008,P.B.Dixon-D.J.Starling-A.N.Jordan-J.C.Howell-2009,M.Iinuma-Y.Suzuki-G.Taguchi-Y.Kadoya-H.F.Hofmann-2011}.
These experiments show that the weak measurement is also very
useful for high-precision measurements.
For example, Hosten and Kwiat~\cite{O.Hosten-P.Kwiat-2008} used
the weak measurement to observe a tiny spin Hall effect in
light; Dixon et
al.~\cite{P.B.Dixon-D.J.Starling-A.N.Jordan-J.C.Howell-2009,D.J.Starling-P.B.Dixon-A.N.Jordan-J.C.Howell-2009}
(DSJH) used the weak measurement to detect very small transverse
beam deflections.
The original AAV work~\cite{Y.Aharonov-D.Z.Albert-L.Vaidman-1988} also includes the
proposal of the application to the sequence of the Stern-Gerlach
experiments for spin-1/2 particles.
They claim that we can observe the spin of particles as a
larger value than the range of its eigenvalues.
This is called ``weak-value amplification''.
The above high-precision measurements using the weak measurement
are due to the effect of this weak-value amplification.


Weak measurements are based on von-Neumann's measurement
theory~\cite{J.von-Neumann-1932} in which the total system
consists of the system to be measured and a detector to measure
the system.
Further, we specify the initial state (pre-selection) and the
final state (post-selection) of the system. 
AAV also proposed the situation of the measurement, in which the 
initial variance in the momentum conjugate to the pointer
variable of the detector is so small that the interaction
between the system and the detector is very
weak~\cite{Aharonov-Vaidman-2007-update-review}. 
Because of this weakness, the measurement proposed by AAV is
called ``weak measurement''.
In the {\it linear-order} of the interaction between the system
and detector, the outcome of the weak measurements is so-called
``weak value''.
The weak-value amplification is essentially due to the fact
that the weak value of an quantum observable may become larger
than eigenvalues of this observable when the pre- and the
post-selection is nearly orthogonal.
Due to the weakness of the interaction between the system and
the detector, the measurement by a single ensemble is imprecise.
However, as noted by Aharonov and
Vaidman~\cite{Aharonov-Vaidman-2007-update-review}, the
measurement become precise by a factor $\sqrt{N}$ through
performing large $N$ ensemble experiments.


Measurements of arbitrary strength beyond the linear-order
interaction has been first discussed by Aharonov and
Botero~\cite{Y.Aharonov-A.Botero-2005} in the context of the
framework called ``Quantum average of weak value.''
In this framework, the strong measurement of a pointer variable
can be regarded as quantum superpositions of weak measurements.
They applied their framework to a specific case of a spin
measurement.
Furthermore, {\it all-order} effects of the unitary evolution
due to the von-Neumann interaction between the system and the
detector are also investigated by investigated by Di Lorenzo and
Euges~\cite{A.D.Lorenzo-J.C.Egues-2008} in AAV setup to clarify
the detector dynamics in weak masurements.


More recently, Wu and Li~\cite{S.Wu-Y.Li-2010} proposed the general 
formulation of the weak measurement which includes 
all-order effects of the unitary evolution due to the
von-Neumann interaction between the system and the detector.
Through this formulation, they took some higher-order effects
into account when they computed the shift of pointer variables
and pointed out that there is a overlap of the pre- and the
post-selection at which the outcome of the weak measurement have
the maximal amplification.
However, since they did not take all higher-order effects into 
account, their claim on the maximal amplification is weak.


In this paper, we carry out the all-order evaluation of some
expectation values of pointer variables after the post-selection
based on the formulation proposed by Wu and
Li~\cite{S.Wu-Y.Li-2010}. 
Although the all-order evaluations of the expectation values in
general weak measurement are difficult, these evaluations are
possible if we concentrate only on the weak measurements for an
operator ${\bf A}$ of the system which satisfies the property 
${\bf A}^{2}=1$.
Choosing the initial state of the detector as a zero mean-value
Gaussian state, we derive some formulae of the expectation
values and probability densities for the detector after the
post-selection without any approximation.
Through these formulae, we discuss the maximal amplification
which suggested by Wu and Li.


Although our consideration is restricted only to the case of the
weak measurement for an operator ${\bf A}$ which satisfies the 
property ${\bf A}^{2}=1$, this case includes many experimental
setups.
For example, the weak measurement of the spins of spin-1/2
particles, which was originally proposed by
AAV~\cite{Y.Aharonov-D.Z.Albert-L.Vaidman-1988}, is included
since the Pauli spin matrices satisfy the property 
${\bf A}^{2}=1$.
The experiment by Hosten-Kwiat~\cite{O.Hosten-P.Kwiat-2008} and
the experimental setup by DSJH~\cite{P.B.Dixon-D.J.Starling-A.N.Jordan-J.C.Howell-2009} are
also included in our case, though there are some additional
modification in their actual experimental setups. 
Thus, our consideration will be applicable to many
experimental setups.
Therefore, it is worthwhile to research the weak measurements
for an operator ${\bf A}$ which satisfies the property 
${\bf A}^{2}=1$.


Furthermore, we note that some experiments of weak measurement
for an operator ${\bf A}$ which satisfies the property 
${\bf A}^{2}=1$ are classified into two types: one is the weak
measurements with a real weak value; and the other is those with 
a weak value of pure imaginary.
A typical example of the weak measurement with a real weak value
is the experimental setup of a spin-1/2 particle proposed by
AAV~\cite{Y.Aharonov-D.Z.Albert-L.Vaidman-1988}. 
On the other hand, a typical example of the weak measurement
with a weak value of pure imaginary is the DSJH
experiment~\cite{P.B.Dixon-D.J.Starling-A.N.Jordan-J.C.Howell-2009}. 
We apply our results of all-order evaluations to these two
specific experimental setups.
Then, we discuss the optimizations of the expectation value of
the pointer variable of the detector (i.e., the signal 
optimization) and the optimization of the signal to noise
ratio (SNR). 
Through these applications, we concretely discuss the maximum 
amplification in the weak measurements.


Organization of this paper is as follows:
In Sec.~\ref{sec:Wu-Li_Formalism}, we briefly review the general
formulation proposed by Wu and Li.
In Sec.~\ref{sec:All-order_evaluation_of_WM_for_A2is1}, we
summarize the formulae for some expectation values and
probability densities which are derived from all-order
evaluations through Wu-Li formulation.
In Sec.~\ref{sec:Application_to_AAV_setup}, the application of
our formulae to AAV setup is discussed.
In Sec.~\ref{sec:Application_to_DSJH_setup}, we discuss the
application of our formulae to DSJH setup, though the
experimental setup in this paper is a simpler version of the
original DSJH setup.
Final section (Sec.~\ref{sec:Summary_Discussion}) is devoted to
the summary.


Throughout this paper, we use the natural unit $\hbar=1$.


\section{Wu-Li Formalism}
\label{sec:Wu-Li_Formalism}


Here, we review the description of weak measurements proposed
by Wu and Li~\cite{S.Wu-Y.Li-2010}.
In Sec.~\ref{sec:Wu-Li_Formalism-general}, we first review the
general framework of the weak measurement following
Ref.~\cite{S.Wu-Y.Li-2010}.
To carry out the analyses, we must treat two cases separately
for a technical reason.
One is the case where the initial and the final states of the
system is not orthogonal, which is described in
Sec.~\ref{sec:Wu-Li_Formalism-non-orthogonal}. 
The other is the case where the initial and the final states of
the system is orthogonal, which is described in
Sec.\ref{sec:Wu-Li_Formalism-orthogonal}.


\subsection{General framework}
\label{sec:Wu-Li_Formalism-general}


The total system we consider here is described by the density
matrix $\rho=\rho_{s}\otimes\rho_{d}$.
$\rho_{s}$ is the density matrix of the ``system'' which is a
quantum system and we measure an observable ${\bf A}$ associated
with this system.
$\rho_{d}$ is the density matrix of the ``detector'' which
interacts with the system through the von-Neumann interaction  
\begin{eqnarray}
  \label{eq:von-Neumann-interaction}
  {\cal H} = g\delta(t-t_{0}){\bf A}\otimes p.
\end{eqnarray}
Here, $p$ is the conjugate momentum to the pointer variable
$q$ of the detector, i.e., $[q,p]=i$.
In the usual von-Neumann interaction (strong interaction), the
eigenvalues of ${\bf A}$ appear in the pointer variable
$q$~\cite{J.von-Neumann-1932}.
Using these three elements, the weak measurement is carried out 
through the sequence of four
measurements~\cite{Y.Aharonov-D.Z.Albert-L.Vaidman-1988}.
First three processes of these four measurements are called
``pre-selection'', ``weak interaction'', ``post-selection''.
The final one is the measurement of the detector pointer
variable through any type of the measurement in quantum
mechanics.


First, we prepare the initial state $\rho_{s}$ of the system
through the projection measurement at $t<t_{0}$, which is called 
``pre-selection''.
We also prepare the initial state of the detector
$\rho_{d}$.
After this pre-selection, the system and the detector interact
with each other through the interaction Hamiltonian
(\ref{eq:von-Neumann-interaction}). 
The time evolution through this interaction is described by the
evolution operator ${\cal U} = e^{-ig{\bf A}p}$ and the total
density matrix $\rho$ evolves as 
\begin{eqnarray}
  \label{eq:evol-of-total-density-matrix}
  \rho' = {\cal U}\rho{\cal U}^{\dagger}
  =
  \rho
  + 
  \sum_{n=1}^{+\infty} \frac{(-ig)^{n}}{n!} {\bf ad}^{n}p{\bf A}\circ\rho, 
\end{eqnarray}
where ${\bf ad}^{n}$ for arbitrary operators $\Omega$ and 
$\Theta$ is recursively defined as 
\begin{eqnarray}
  \label{eq:adnprho-def-1}
  {\bf ad}^{1}\Omega\circ\Theta
  &:=&
  {\bf ad}\Omega\circ\Theta
  =
  \left[\Omega,\Theta\right]
  , \\
  {\bf ad}^{n}\Omega\circ\Theta
  &:=&
  {\bf ad}\Omega\circ({\bf ad}^{n-1}\Omega\circ\Theta)
  \nonumber\\
  &=&
  \left[\Omega,{\bf ad}^{n-1}\Omega\circ\Theta\right].
  \label{eq:adnprho-def-n}
\end{eqnarray}
The prime in Eq.~(\ref{eq:evol-of-total-density-matrix}) denotes
the operator after the interaction
(\ref{eq:von-Neumann-interaction}).
The density matrix of the system after this interaction is given
by 
\begin{eqnarray}
  \label{eq:system-density-metric-after-weak-int}
  \rho'_{s}
  =
  \TR_{d}\rho'
  =
  \rho_{s} 
  + \sum_{n=1}^{+\infty} \frac{(-ig)^{n}}{n!}\langle p^{n}\rangle
  {\bf ad}^{n}{\bf A}\circ\rho_{s}
  ,
\end{eqnarray}
where $\TR_{d}$ means taking the trace of the detector density
matrix and $\langle p^{n}\rangle:=\TR_{d}(p^{n}\rho_{d})$.
Equation (\ref{eq:system-density-metric-after-weak-int}) implies
that the density matrix of the system hardly changes through the
interaction with the detector if $g\sup\{\langle
p^{n}\rangle^{1/n},n\in\NN\}\ll 1$.
Roughly speaking, this condition is regarded as 
$g\Delta p\ll 1$, where $(\Delta p)^{2}$ is the variance in $p$,
and is interpreted that the interaction between the system and
the detector in the measurement is ``weak interaction''. 
After this interaction, we restrict the final state of the
system by the projection operator $\Pi_{f}$:
$\rho'\rightarrow\rho'\Pi_{f}$. 
This restriction is called ``post-selection''.
The density matrix of the detector after the post-selection is
given by 
\begin{eqnarray}
  \label{eq:detector-density-metric-after-PS}
  \rho_{d}'
  =
  \frac{
    \TR_{s}\rho'\Pi_{f}
  }{
    \TR\rho'\Pi_{f}
  }
  ,
\end{eqnarray}
where $\TR_{s}$ ($\TR$) means taking the trace of the system
density matrix (the total density matrix).


Although $\rho_{s}$ and $\rho_{d}$ may describe mixed states of
the system and the detector, we restrict our attention to pure
states as the initial density matrices $\rho_{s}$ and
$\rho_{d}$. 
We denote these initial density matrices as
$\rho_{s}=|\psi_{i}\rangle\langle\psi_{i}|$ and
$\rho_{d}=|\phi\rangle\langle\phi|$. 
Further, we also denote the projection operator for the
post-selection by $\Pi_{f}:=|\psi_{f}\rangle\langle\psi_{f}|$. 
In this case, the normalization factor ($\TR\rho'\Pi_{f}$) of
the density matrix of the detector after the post-selection is
given by 
\begin{eqnarray}
  \label{eq:normalization-detector-density-matrix-after-PS}
  \TR\rho'\Pi_{f}
  &=&
  \left|\langle\psi_{f}|\psi_{i}\rangle\right|^{2}
  +
  \sum_{n=1}^{+\infty} \frac{(-ig)^{n}\langle p^{n}\rangle}{n!}
  \sum_{k=0}^{n}(-1)^{k} {}_{n}C_{k}
  \nonumber\\
  && \quad\quad\quad\quad\quad
  \times
  \langle\psi_{f}|{\bf A}^{n-k}|\psi_{i}\rangle
  \langle\psi_{i}|{\bf A}^{k}|\psi_{f}\rangle,
\end{eqnarray}
where $\langle p^{n}\rangle = \langle\phi|p^{n}|\phi\rangle$ and
${}_{n}C_{k}$ is the binomial coefficient.


To carry out the further analyses, the factor 
$\left|\langle\psi_{f}|\psi_{i}\rangle\right|$ plays an
important role and separate treatments are required according to
the fact whether
$\left|\langle\psi_{f}|\psi_{i}\rangle\right|=0$ or not.


\subsection{Non-orthogonal weak measurement
  $\left|\langle\psi_{f}|\psi_{i}\rangle\right|\neq 0$}
\label{sec:Wu-Li_Formalism-non-orthogonal}


Here, we consider the case where the pre- and post-selection are
not orthogonal, i.e., 
$\left|\langle\psi_{f}|\psi_{i}\rangle\right|\neq 0$.
In this case, the trace of the post-selected density matrix and
the density matrix $\rho_{d}'$ after the post-selection are
given by 
\begin{eqnarray}
  \label{eq:total-trace-of-PS-density-matrix}
  \TR\rho'\Pi_{f}
  &=:&
  \left|\langle\psi_{f}|\psi_{i}\rangle\right|^{2}
  {\cal Z}
  ,
  \\
  {\cal Z}
  &=&
  1
  +
  \sum_{n=1}^{+\infty} \frac{(-ig)^{n}\langle p^{n}\rangle}{n!}
  \nonumber\\
  && \quad\quad
  \times
  \sum_{k=0}^{n} (-1)^{k} {}_{n}C_{k}
  \langle {\bf A}^{n-k}\rangle_{w} \langle {\bf A}^{k}\rangle_{w}^{*}
  , \nonumber\\
  \label{eq:total-trace-of-PS-density-matrix-Z}
  \\
  {\cal Z}\rho_{d}'
  &=&
  \rho_{d}
  +
  \sum_{n=1}^{+\infty} \frac{(-ig)^{n}}{n!} 
  \sum_{k=0}^{n}(-1)^{k} {}_{n}C_{k}
  \nonumber\\
  && \quad\quad\quad
  \times
  \langle{\bf A}^{n-k}\rangle_{w} 
  \langle{\bf A}^{k}\rangle_{w}^{*} 
  p^{n-k}\rho_{d}p^{k}
  ,
  \nonumber\\
  \label{eq:detector-density-matrix-after-PS}
\end{eqnarray}
where $\langle\cdot\rangle_{w} :=
\langle\psi_{f}|\cdot|\psi_{i}\rangle/\langle\psi_{f}|\psi_{i}\rangle$.


When the wave function $\langle p|\phi\rangle$ is even in $p$,
i.e., $\langle p^{n}\rangle=0$ for odd $n$, Wu and Li derived the
formulae of the shifts in $q$ and $p$ as
\begin{eqnarray}
  \label{eq:S.Wu-Y.Li-2010-19}
  \delta q
  &=&
  \frac{
    g\Re{\bf A}_{w} + g \Im{\bf A}_{w} \langle\left\{q,p\right\}\rangle
  }{
    1 + g^{2}\langle p^{2}\rangle\left(
      \left|{\bf A}_{w}\right|^{2}
      -
      \Re\langle{\bf A}^{2}\rangle_{w}
    \right)
  }
  , \\
  \label{eq:S.Wu-Y.Li-2010-20}
  \delta p
  &=&
  \frac{
    2 g\Im{\bf A}_{w} \langle p^{2}\rangle
  }{
    1 + g^{2} \langle p^{2}\rangle \left(
      \left|{\bf A}_{w}\right|^{2}
      -
      \Re\langle{\bf A}^{2}\rangle_{w}
    \right)
  }
  ,
\end{eqnarray}
where 
$\delta q:=\TR(q\rho_{d}')-\TR(q\rho_{d})$ and 
$\delta p:=\TR(p\rho_{d}')-\TR(p\rho_{d})$.
In their derivation, they neglect terms of $O(g^{3})$ in the
numerators and the denominators, but they do not expand the
total expressions (\ref{eq:S.Wu-Y.Li-2010-19}) and 
(\ref{eq:S.Wu-Y.Li-2010-20}) in form of the power series of $g$.
Although these treatments of $\delta q$ and $\delta p$ might be
regarded as some renormalization technique, it is also true that
the expressions (\ref{eq:S.Wu-Y.Li-2010-19}) and
(\ref{eq:S.Wu-Y.Li-2010-20}) include only partial effects of
higher order of $g$.


As pointed out by
AAV~\cite{Y.Aharonov-D.Z.Albert-L.Vaidman-1988}, weak values may 
become very large in the limit
$\left|\langle\psi_{f}|\psi_{i}\rangle\right|\rightarrow 0$
($\neq 0$).
At the order of $O(g)$, the shifts (\ref{eq:S.Wu-Y.Li-2010-19})
and (\ref{eq:S.Wu-Y.Li-2010-20}) are proportional to the weak
value $\langle {\bf A}\rangle_{w}$~\cite{R.Jozsa-2007}. 
This implies that the shifts (\ref{eq:S.Wu-Y.Li-2010-19}) and
(\ref{eq:S.Wu-Y.Li-2010-20}) of order $O(g)$ may diverge in the 
limit $\left|\langle\psi_{f}|\psi_{i}\rangle\right|\rightarrow 0$.  
This is the essence of the weak value amplification.
However, from the total expressions of
Eqs.~(\ref{eq:S.Wu-Y.Li-2010-19}) and
(\ref{eq:S.Wu-Y.Li-2010-20}), Wu and Li suggested that, in the
limit $\left|\langle\psi_{f}|\psi_{i}\rangle\right|\rightarrow 0$, 
these shifts decrease rapidly when 
$\left|{\bf A}_{w}\right|^{2}$ become comparable with
$(g^{2}\langle p^{2}\rangle)^{-1}$.
This arguments implies that, for a fixed 
$g^{2}\langle p^{2}\rangle$, there may exist a maximum shift of
a pointer quantity, and an optimal overlap
$\left|\langle\psi_{f}|\psi_{i}\rangle\right|$ to achieve the
maximum shift.
We call this overlap as the {\it optimal pre-selection} (or 
{\it optimal post-selection}).


Although Wu and Li claim is weak in the sense that they did not
take all higher-order effects into account, in this paper, we
show that their claim on the optimal pre-selection is essentially
correct through the all-order evaluation of weak measurements
for an operator ${\bf A}$ which satisfies the property 
${\bf A}^{2}=1$.


\subsection{Orthogonal weak measurement $\left|\langle\psi_{f}|\psi_{i}\rangle\right|= 0$}
\label{sec:Wu-Li_Formalism-orthogonal}


Next, we consider the orthogonal case where
$\left|\langle\psi_{f}|\psi_{i}\rangle\right|= 0$. 
In this case, the original formalism of the weak measurement
fails and the weak values are not defined.
This is easily seen from the fact that the normalization factor
${\cal Z}$ defined by
Eq.~(\ref{eq:total-trace-of-PS-density-matrix}) is ill-defined. 
However, instead of ${\cal Z}$, Wu and Li defined ${\cal Z}_{o}$ 
by 
\begin{eqnarray}
  \label{eq:S.Wu-Y.Li-2010-22-0}
  \TR\rho'\Pi_{f}
  =:
  g^{2}\langle p^{2}\rangle
  \left|\langle\psi_{f}|{\bf A}|\psi_{i}\rangle\right|^{2}
  {\cal Z}_{o} 
  ,
\end{eqnarray}
\begin{eqnarray}
  {\cal Z}_{o} &=& 1 + \sum_{n=1}^{+\infty} \frac{(-ig)^{n}}{n!}
  \frac{\langle p^{n+2}\rangle}{\langle p^{2}\rangle}
  \nonumber\\
  && \quad\quad
  \times
  \sum_{k=0}^{n}(-1)^{k} {}_{n}\!C_{k}
  \langle{\bf A}^{n-k}\rangle_{ow} 
  \langle{\bf A}^{k}\rangle_{ow}^{*} 
  ,
  \nonumber\\
  \label{eq:S.Wu-Y.Li-2010-22}
\end{eqnarray}
where
\begin{eqnarray}
  \label{eq:S.Wu-Y.Li-2010-23}
  \langle{\bf A}^{n}\rangle_{ow}
  :=
  \frac{
    \langle\psi_{f}|{\bf A}^{n+1}|\psi_{i}\rangle
  }{
    \langle\psi_{f}|{\bf A}(n+1)|\psi_{i}\rangle
  }
  .
\end{eqnarray}
Wu and Li called $\langle{\bf A}^{n}\rangle_{ow}$ defined by
Eq.~(\ref{eq:S.Wu-Y.Li-2010-23}) as orthogonal weak values. 
The density matrix of the detector after the post-selection is
given by 
\begin{eqnarray}
  \label{eq:S.Wu-Y.Li-2010-24}
  {\cal Z}_{o}\langle p^{2}\rangle\rho_{d}'
  &=& 
  p\rho_{d}p
  \nonumber\\
  &&
  +
  \sum_{n=1}^{+\infty} \frac{(-ig)^{n}}{n!}
  \sum_{k=0}^{n}(-1)^{k} {}_{n}\!C_{k}
  \langle{\bf A}^{n-k}\rangle_{ow}
  \langle{\bf A}^{k}\rangle_{ow}^{*}
  \nonumber\\
  && \quad\quad\quad
  \times
  p^{n-k+1}\rho_{d}p^{k+1}
  .
\end{eqnarray}
From this expression (\ref{eq:S.Wu-Y.Li-2010-24}), Wu and Li
claim that the orthogonal weak values
(\ref{eq:S.Wu-Y.Li-2010-23}) play the similar role to the
original weak values in non-orthogonal case.


\section{All-order evaluation of weak measurements for an
  operator ${\bf A}$ which satisfies ${\bf A}^{2}=1$}
\label{sec:All-order_evaluation_of_WM_for_A2is1}


Here, we evaluate the density matrix of the detector after the
post-selection and some expectation values in the case for an
operator ${\bf A}$ which satisfies the property ${\bf A}^{2}=1$ 
based on the Wu-Li formalism.
In addition to the restriction of our consideration to the
simple operator case, in this section, we assume that the
initial state $\rho_{d}=|\phi\rangle\langle\phi|$ of the
detector is zero mean-value Gaussian, i.e.,
\begin{eqnarray}
  \label{eq:instate-of-detector-is-Gaussian}
  \langle p|\phi\rangle
  =
  \left(
    \frac{1}{2\pi\langle p^{2}\rangle}
  \right)^{1/4}
  \exp\left[
    - \frac{p^{2}}{4\langle p^{2}\rangle}
  \right]
  .
\end{eqnarray}
From this initial state of the detector, we can easily derive
the properties of the initial state:
\begin{eqnarray}
  \label{eq:instate-p-moments-in-Gaussian}
  \langle p^{2n+1}\rangle = 0, \quad
  \langle p^{2n}\rangle = (2n-1)!! \langle p^{2}\rangle^{n}.
\end{eqnarray}


As reviewed in the last section \ref{sec:Wu-Li_Formalism},
according to the norm
$\left|\langle\psi_{f}|\psi_{i}\rangle\right|^{2}$, we have to
treat the density matrix in different way.
Therefore, we treat a non-orthogonal weak measurement and an 
orthogonal one, separately.


\subsection{Non-orthogonal weak measurement
  $\left|\langle\psi_{f}|\psi_{i}\rangle\right|\neq 0$}
\label{sec:A2is1case-non-orthogonal}


When the initial state of the detector is zero mean-value
Gaussian (\ref{eq:instate-of-detector-is-Gaussian}), the moments 
of $p$ are given by Eqs.~(\ref{eq:instate-p-moments-in-Gaussian}).
In this case, the normalization ${\cal Z}$
[Eq.~(\ref{eq:kouchan-A2is1-9})] is given by  
\begin{eqnarray}
  {\cal Z}
  =
  1
  +
  \frac{1}{2}
  \left(
    1
    -
    \left|\langle{\bf A}\rangle_{w}\right|^{2}
  \right)
  \left(
    e^{-s} - 1
  \right)
  ,
  \label{eq:kouchan-A2is1-10-2}
\end{eqnarray}
where $s$ is a parameter defined by 
\begin{eqnarray}
  \label{eq:coupling-parameter-def}
  s := 2g^{2}\langle p^{2}\rangle.
\end{eqnarray}
Similar calculations lead the expectation values of $p$ and $q$
after the post-selection
\begin{eqnarray}
  \frac{\langle q\rangle'}{g}
  &=&
  \frac{
    \Re\langle{\bf A}\rangle_{w}
  }{
    {\cal Z}
  }
  ,
  \label{eq:kouchan-A2is1-52}
  \\
  g \langle p\rangle'
  &=&
  \frac{
    s e^{-s} \Im\langle{\bf A}\rangle_{w}
  }{
    {\cal Z}
  }
  .
  \label{eq:kouchan-A2is1-53}
\end{eqnarray}
[Here, we denotes the expectation value of $*$ for the detector
after the post-selection by $\langle *\rangle'$.
Fluctuations 
$\Delta q:=\sqrt{\langle(q-\langle q\rangle')^{2}\rangle'}$ 
and $\Delta p:=\sqrt{\langle(p-\langle p\rangle')^{2}\rangle'}$
in $p$ and $q$ after the post-selection are given by 
\begin{eqnarray}
  \frac{(\Delta q)^{2}}{g^{2}}
  &=&
  \frac{1}{2s}
  + \frac{1}{2{\cal Z}} \left(
    1 + \left|\langle{\bf A}\rangle_{w}\right|^{2}
  \right)
  \nonumber\\
  &&
  -
  \frac{
    \left(\Re\langle{\bf A}\rangle_{w}\right)^{2}
  }{
    {\cal Z}^{2}
  }
  ,
  \label{eq:kouchan-A2is1-94}
  \\
  g^{2}(\Delta p)^{2}
  &=&
  \frac{s}{2}
  -
  \frac{s^{2}e^{-s}}{2{\cal Z}} \left(
    1 - \left|\langle{\bf A}\rangle_{w}\right|^{2}
  \right)
  \nonumber\\
  &&
  -
  \frac{s^{2}e^{-2s}(\Im\langle{\bf A}\rangle_{w})^{2}}{{\cal Z}^{2}}
  .
  \label{eq:kouchan-A2is1-95}
\end{eqnarray}


Further, the probability densities in $p$-space and $q$-space are
given by 
\begin{eqnarray}
  \langle p|\rho_{d}'|p\rangle
  &=&
  \left[
    2
    +
    \left(
      1 - \left|\langle{\bf A}\rangle_{w}\right|^{2}
    \right)
    \left(
      \cos(2gp) - 1
    \right)
  \right.
  \nonumber\\
  && \;\;
  \left.
    + 2 \Im\langle{\bf A}\rangle_{w} \sin(2gp)
    \frac{}{}
  \right]
  \frac{
    \langle p|\rho_{d}|p\rangle
  }{
    2 {\cal Z}
  }
  ,
  \label{eq:kouchan-A2is1-101}
  \\
  \langle q|\rho_{d}'|q\rangle
  &=&
  \left[\frac{}{}
    1
    - \left|\langle{\bf A}\rangle_{w}\right|^{2}
  \right.
  \nonumber\\
  && \;\;
  \left.
    + \left(
      1
      + \left|\langle{\bf A}\rangle_{w}\right|^{2}
    \right) \cosh\left(\frac{2sq}{g}\right)
  \right.
  \nonumber\\
  && \;\;
  \left.
    + 2 \Re\langle{\bf A}\rangle_{w} \sinh\left(\frac{2sq}{g}\right)
  \right]
  \nonumber\\
  && \quad\quad\quad\quad
  \times
  \frac{
    e^{-s} \langle q|\rho_{d}|q\rangle
  }{
    2 {\cal Z}
  }
  .
  \label{eq:kouchan-A2is1-141}
\end{eqnarray}
where $\langle p|\rho_{d}|p\rangle$ and 
$\langle q|\rho_{d}|q\rangle$ are Gaussian initial probability
densities 
\begin{eqnarray}
  \label{eq:initial-Gaussian-prob-density-p-space}
  \langle p|\rho_{d}|p\rangle
  &=&
  \sqrt{\frac{g^{2}}{\pi s}}
  \exp\left[- \frac{(gp)^{2}}{s}\right]
  ,
  \\
  \label{eq:initial-Gaussian-prob-density-q-space}
  \langle q|\rho_{d}|q\rangle
  &=&
  \sqrt{\frac{s}{\pi g^{2}}}
  \exp\left[
    - s \left(\frac{q}{g}\right)^{2}
  \right]
  .
\end{eqnarray}
The derivation of Eq.~(\ref{eq:kouchan-A2is1-141}) is explained
in Appendix \ref{sec:Derivations_of_formulae_non_orthogonal}.


Here, we note that the parameter $s$ defined in
Eq.~(\ref{eq:coupling-parameter-def}) is a measure of the
strength of the interaction.
Usually, it is said that the interaction between the system and
the detector is weak if the coupling constant $g$ is very small.
On the other hand, in the weak
measurement~\cite{Aharonov-Vaidman-2007-update-review}, it is
said that the interaction between the system and the detector is
weak if the initial variance of the pointer variable $q$ is very
large, i.e., the variance in the conjugate momentum $p$ is very
small.
These two concepts of the ``weakness'' of the measurement are
automatically represented by the single non-dimensional
parameter $s$.
We call $s$ as the coupling parameter, and say that the
interaction between the system and the detector is weak if 
$s\ll 1$ and strong if $s\gg 1$.


We have to emphasize that our formulae shown here are
the results from the all-order evaluation of $s$ and valid not
only in the weak measurement regime $s\ll 1$ but also in the 
strong measurement regime $s\gg 1$.
The results coincide with those of the measurement in the strong
regime.
This situation can be observed through the specific experimental
setups discussed in Sec.~\ref{sec:Application_to_AAV_setup}.


Finally, we note that probability distributions for weak
measurements (both in the strong and weak regime) are first
discussed by Aharonov and Botero~\cite{Y.Aharonov-A.Botero-2005}
in the context of the framework called ``Quantum averages of
weak value'' as mentioned in Sec.~\ref{sec:intro}.
Of course, our formulae of the probability distribution shown in
this paper are not general because we concentrate only on the
case of an operator ${\bf A}$ which satisfies the property 
${\bf A}^{2}=1$.
However, as emphasize in Sec.~\ref{sec:intro}, many experimental
setups are included in this special case and we have derived
explicit simple analytic formulae for this special case.
This is one of main points of this paper.


\subsection{Orthogonal weak measurement
  $\left|\langle\psi_{f}|\psi_{i}\rangle\right|= 0$} 
\label{sec:A2is1case-orthogonal}


Now, we consider the orthogonal case where the pre-selected state
and the post-selected state are orthogonal to each other, i.e.,
$\langle\psi_{f}|\psi_{i}\rangle = 0$. 
As reviewed in Sec.~\ref{sec:Wu-Li_Formalism-orthogonal}, the
density matrix of the detector after the post-selection is given
by Eq.~(\ref{eq:S.Wu-Y.Li-2010-24}).
In the case of the weak measurements for the operator ${\bf A}$
with the property ${\bf A}^{2}=1$, the orthogonal weak values
(\ref{eq:S.Wu-Y.Li-2010-23}) are given by 
\begin{eqnarray}
  \label{eq:kouchan-A2is1-orthogonal-14}
  \langle{\bf A}^{n}\rangle_{ow}
  = \left\{
    \begin{array}{ccc}
      \displaystyle
      \frac{1}{n+1} & \mbox{for} & n \;\; \mbox{is even}, \\
                  0 & \mbox{for} & n \;\; \mbox{is odd}.
    \end{array}
  \right.
\end{eqnarray}
This expression implies that no information of ${\bf A}$ appears
in the orthogonal weak measurement for an operator ${\bf A}$
with the property ${\bf A}^{2}=1$.


Through the Gaussian initial state
(\ref{eq:instate-of-detector-is-Gaussian}) of the detector with
the properties (\ref{eq:instate-p-moments-in-Gaussian}), the
normalization constant ${\cal Z}_{o}$ defined by
Eq.~(\ref{eq:S.Wu-Y.Li-2010-22}) is given by 
\begin{eqnarray}
  {\cal Z}_{o}
  &=&
  \frac{4}{s} \left(
    1
    - e^{-s}
    - \frac{3}{4} s
  \right)
  ,
  \label{eq:kouchan-A2is1-orthogonal-17-1}
\end{eqnarray}
where the coupling parameter $s$ is defined in
Eq.~(\ref{eq:coupling-parameter-def}).
The density matrix (\ref{eq:S.Wu-Y.Li-2010-24}) of the detector
after the post-selection is given by 
\begin{eqnarray}
  {\cal Z}_{o} \langle p^{2}\rangle \rho_{d}'
  &=&
  p\rho_{d}p
  +
  \sum_{n=1}^{+\infty} \frac{(-ig)^{2n}}{(2n+2)!}
  \sum_{k=0}^{n} {}_{2n+2}C_{2k+1}
  \nonumber\\
  && \quad\quad\quad\quad\quad
  \times
  p^{2n-2k+1}\rho_{d}p^{2k+1}
  .
  \label{eq:kouchan-A2is1-orthogonal-2.3.2}
\end{eqnarray}


In the case where the initial state of the detector is zero
mean-value Gaussian (\ref{eq:instate-of-detector-is-Gaussian}),
the expectation value of $p$ after the post-selection, which is
evaluated in Appendix
\ref{sec:Derivations_of_formulae_orthogonal}, is trivial,
\begin{eqnarray}
  \langle p\rangle' = 0,
  \label{eq:kouchan-A2is1-orthogonal-2.4.2}
\end{eqnarray}
due to the properties (\ref{eq:instate-p-moments-in-Gaussian}).
Further, in Appendix
\ref{sec:Derivations_of_formulae_orthogonal}, we also show that
the expectation value of $q$
[Eq.~(\ref{eq:kouchan-A2is1-orthogonal-2.4.3})] after the
post-selection also yields a trivial result
\begin{eqnarray}
  \langle q\rangle' = 0.
  \label{eq:kouchan-A2is1-orthogonal-2.4.6}
\end{eqnarray}


As shown in Appendix
\ref{sec:Derivations_of_formulae_orthogonal}, the fluctuations
$\Delta p$ and $\Delta q$ in $p$ and $q$ of the detector after 
the post-selection are given by 
\begin{eqnarray}
  g^{2}(\Delta p)^{2} 
  &=&
  \frac{s}{2}
  \frac{
    1 + (2 s - 1)e^{-s}
  }{
    4 - 3 s - 4 e^{-s}
  }
  .
  \label{eq:kouchan-A2is1-orthogonal-2.4.11}
  \\
  \frac{(\Delta q)^{2}}{g^{2}}
  &=&
  \frac{1}{2s}
  \frac{1 - e^{-s} + 4s}{4 - 4e^{-s} - 3s}
  .
  \label{eq:kouchan-A2is1-orthogonal-2.4.15}
\end{eqnarray}


The probability densities in $p$-space and in $q$-space are
given by
\begin{eqnarray}
  \langle p|\rho_{d}'|p\rangle
  &=&
  \frac{1 - \cos(2gp)}{ 2 \left( 4 - 4 e^{-s} - 3 s \right) }
  \langle p|\rho_{d}|p\rangle
  ,
  \label{eq:kouchan-A2is1-orthogonal-2.6.4}
  \\
  \langle q|\rho_{d}'|q\rangle
  &=&
  \frac{
    2e^{-s}
    \sinh^{2}\left(sq/g\right)
  }{
    4 - 4e^{-s} - 3 s
  }
  \langle q|\rho_{d}|q\rangle
  ,
  \label{eq:kouchan-A2is1-orthogonal-2.7.6}
\end{eqnarray}
respectively.
Here, $\langle p|\rho_{d}|p\rangle$ and 
$\langle q|\rho_{d}|q\rangle$ are the Gaussian initial
probability densities
(\ref{eq:initial-Gaussian-prob-density-p-space}) and
(\ref{eq:initial-Gaussian-prob-density-q-space}), respectively.  
The derivations of these formulae are given in Appendix
\ref{sec:Derivations_of_formulae_orthogonal}.


Thus, both in the non-orthogonal weak measurements
(Sec.~\ref{sec:A2is1case-non-orthogonal}) and the orthogonal
one (Sec.~\ref{sec:A2is1case-orthogonal}), we explicitly
derived the analytical expressions of the expectation values of
$p$ and $q$, fluctuations in $p$ and $q$, and the probability
distributions in $p$-space and in $q$-space for the detector
only under two assumptions, i.e., the operator ${\bf A}$ for the
system satisfies the property ${\bf A}^{2}=1$ and the initial
state of the detector is zero mean-value Gaussian
(\ref{eq:instate-of-detector-is-Gaussian}).


We note that the formulae (\ref{eq:kouchan-A2is1-52}) and
(\ref{eq:kouchan-A2is1-53}) for the expectation values for $p$
and $q$ coincide with Eqs.~(\ref{eq:S.Wu-Y.Li-2010-19}) and
(\ref{eq:S.Wu-Y.Li-2010-20}), respectively, if we ignore the
higher-order terms of than $O(g^{2})$.
In this sense, equations (\ref{eq:kouchan-A2is1-52}) and
(\ref{eq:kouchan-A2is1-53}) are all-order extension of
Eqs.~(\ref{eq:S.Wu-Y.Li-2010-19}) and
(\ref{eq:S.Wu-Y.Li-2010-20}) derived by Wu and
Li~\cite{S.Wu-Y.Li-2010}. 
We also note that the expressions of
Eqs.~(\ref{eq:kouchan-A2is1-52}) and (\ref{eq:kouchan-A2is1-53})
are valid for arbitrary value of the coupling parameter $s$. 
Furthermore, the behaviors of Eqs.~(\ref{eq:kouchan-A2is1-52})
and (\ref{eq:kouchan-A2is1-53}) are qualitatively same as those
of Eqs.~(\ref{eq:S.Wu-Y.Li-2010-19}) and 
(\ref{eq:S.Wu-Y.Li-2010-20}).
Therefore, we may say that the claim on the optimal
post-selection proposed by Wu and Li is essentially correct and
mathematically justified by Eqs.~(\ref{eq:kouchan-A2is1-52}) and
(\ref{eq:kouchan-A2is1-53}).


In the following two sections, we apply the formulae summarized
in this section to two specific experimental setups and examine 
the weak measurement of these two setups in detail.
Since we already showed that the orthogonal weak measurement
yields trivial results in the expectation value of $p$ and $q$,
we concentrate only on the non-orthogonal weak measurement.


\begin{figure*}
  \centering
  \includegraphics[width=\textwidth]{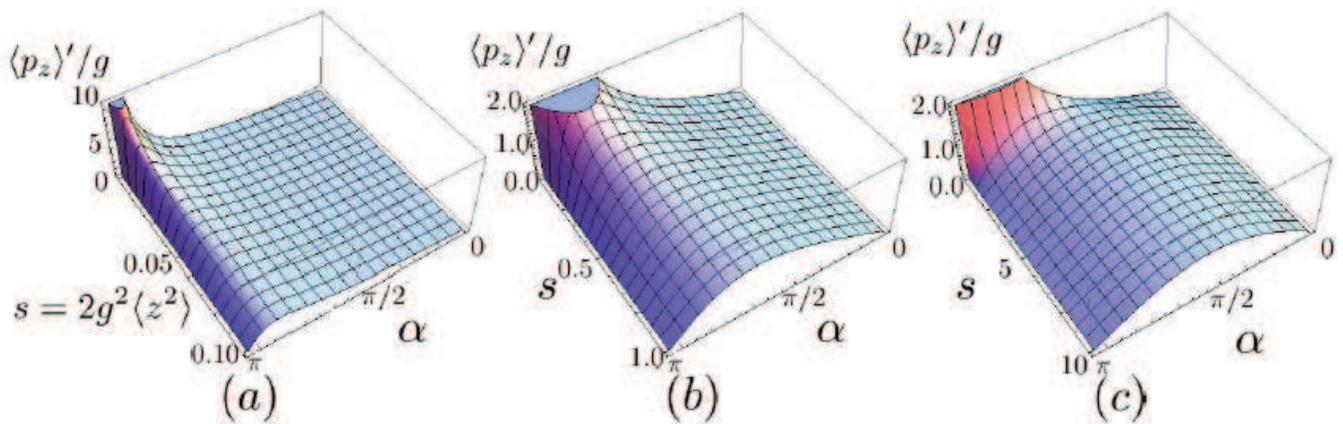}
  \caption{
    $\langle p_{z}\rangle'/g$ is shown as a function of the
    coupling $s=2g^{2}\langle z^{2}\rangle$ and the
    pre-selection angle $\alpha$ in various range of $s$.
    (a) $s\in(0,0.1)$, (b) $s\in(0,1)$, (c) $s\in(0,10)$.
    From these figures, we can see that, for given coupling
    parameter $s$, there is a optimal pre-selection angle
    $\alpha$ such that the expectation value of $p_{z}$ is
    maximized.
  }
  \label{fig:kouchan-AAV-Expect-for-PRA-final}
\end{figure*}


\section{Application to AAV setup} 
\label{sec:Application_to_AAV_setup}


In this section, we apply our formulae derived in 
Sec.~\ref{sec:All-order_evaluation_of_WM_for_A2is1} to the
AAV~\cite{Y.Aharonov-D.Z.Albert-L.Vaidman-1988} setup.
Through this application, we discuss the optimization of the
expectation value of the signal and SNR.


\subsection{Setup of experiment}
\label{sec:AAV_setup_typical}


The experimental setup proposed by
AAV~\cite{Y.Aharonov-D.Z.Albert-L.Vaidman-1988} is the sequence
of three Stern-Gerlach experiments for spin-1/2 particles.


The pre-selected state of the spin-1/2 particle is
\begin{eqnarray}
  \label{eq:AAV-pre-selection}
  |\uparrow_{\xi}\rangle
  =
  \frac{1}{\sqrt{2}} \left(
    \sqrt{1+\sin\alpha}|\uparrow_{z}\rangle
    +
    \sqrt{1-\sin\alpha}|\downarrow_{z}\rangle
  \right)
  ,
\end{eqnarray}
which is an eigenstate
$\sigma_{\xi}|\uparrow_{\xi}\rangle=+|\uparrow_{\xi}\rangle$ of
the operator
$\sigma_{\xi}=\cos\alpha\sigma_{x}+\sin\alpha\sigma_{z}$.


The weak interaction in the weak measurement
is described by the interaction Hamiltonian 
\begin{eqnarray}
  \label{eq:AAV-von-Neumann-interaction}
  {\cal H} = - g z\otimes\sigma_{z}\delta(t-t_{0}), \quad
  g = \mu\left(\frac{\partial B_{z}}{\partial z}\right),
\end{eqnarray}
where $\mu$ is the magnetic moment of the spin-1/2 particle and
$B_{z}$ is the $z$-component of the magnetic field.
The pointer variable in this setup is $p_{z}$ which conjugate to
$z$. 
We note that the operator ${\bf A}$ to be observed is the spin
$z$-component of a spin-1/2 particle through the von-Neumann
interaction (\ref{eq:von-Neumann-interaction}), i.e.,
\begin{eqnarray}
  {\bf A} = \sigma_{z}
  =
  |\uparrow_{z}\rangle\langle\uparrow_{z}|
  -
  |\downarrow_{z}\rangle\langle\downarrow_{z}|
  ,
\end{eqnarray}
which satisfies the property ${\bf A}^{2}=1$.
Then, we may apply our formulae provided in
Sec.~\ref{sec:A2is1case-non-orthogonal}.
Comparing Eq.~(\ref{eq:AAV-von-Neumann-interaction}) with the
interaction Hamiltonian (\ref{eq:von-Neumann-interaction}), we
find the correspondence of variables as
\begin{eqnarray}
  p \rightarrow - z, \quad
  q \rightarrow p_{z}.
\end{eqnarray}


The post-selection in this setup is 
\begin{eqnarray}
  \label{eq:AAV-post-selection}
  |\uparrow_{x}\rangle
  =
  \frac{1}{\sqrt{2}} \left(
    |\uparrow_{z}\rangle + |\downarrow_{z}\rangle
  \right)
\end{eqnarray}
which is an eigenstate
$\sigma_{x}|\uparrow_{x}\rangle=+|\uparrow_{x}\rangle$ of the
$x$-component $\sigma_{x}$ of the spin.


The weak value in this setup is given by 
\begin{eqnarray}
  \label{eq:AAV-weak-value}
  \langle {\bf A}\rangle_{w} =
  \frac{
    \langle\uparrow_{x}|\sigma_{z}|\uparrow_{\xi}\rangle
  }{
    \langle\uparrow_{x}|\uparrow_{\xi}\rangle
  }
  =
  \tan\frac{\alpha}{2}
  ,
\end{eqnarray}
where $\alpha$ is the pre-selection angle.
We note that this weak value (\ref{eq:AAV-weak-value}) is real.


\subsection{All-order expectation values and probability distribution} 
\label{sec:AAV-All-order}


Here, we apply the formulae summarized in
Sec.~\ref{sec:A2is1case-non-orthogonal} to the AAV setup.
The normalization factor ${\cal Z}$
[Eq.~(\ref{eq:kouchan-A2is1-10-2})], is given by  
\begin{eqnarray}
  \label{eq:AAV-normalization}
  {\cal Z}
  =
  \frac{
    1 + e^{-s} \cos\alpha
  }{
    1 + \cos\alpha
  }
  ,
\end{eqnarray}
where $s$ is the coupling parameter [see
Eq.~(\ref{eq:coupling-parameter-def})] defined by 
\begin{eqnarray}
  s := 2 g^{2}\langle z^{2}\rangle.
\end{eqnarray}
Expectation value of $p_{z}$ and $z$ are given by 
\begin{eqnarray}
  \label{eq:AAV-expectation-value}
  \frac{\langle p_{z}\rangle'}{g}
  =
  \frac{\sin\alpha}{1 + e^{-s}\cos\alpha}
  , \quad
  \langle z\rangle' = 0.
\end{eqnarray}
The expectation value of $\langle p_{z}\rangle'/g$ is shown 
as a function of the coupling parameter $s$ and the
pre-selection angle $\alpha$ in
Fig.~\ref{fig:kouchan-AAV-Expect-for-PRA-final}. 
In Fig.~\ref{fig:kouchan-AAV-Expect-for-PRA-final}(a), we can
see a pole at $(s,\alpha)=(0,\pi)$.
Due to this pole, the weak value is amplified as pointed out by
AAV.
In the region $s\gg 1$, the expectation value
(\ref{eq:AAV-expectation-value}) of $\langle p_{z}\rangle'/g$
behaves $\langle p_{z}\rangle'/g\sim\sin\alpha$.
This behavior can be seen in
Fig.~\ref{fig:kouchan-AAV-Expect-for-PRA-final}(c). 
The qualitative difference between
Eqs.~(\ref{eq:S.Wu-Y.Li-2010-19})-(\ref{eq:S.Wu-Y.Li-2010-20})
by Wu-Li and
Eqs.~(\ref{eq:kouchan-A2is1-52})-(\ref{eq:kouchan-A2is1-53}) in
this paper becomes large in the strong region $s\gg 1$.


Fluctuations $\Delta p_{z}$ and $\Delta z$ are given by 
\begin{eqnarray}
  \frac{\left(\Delta p_{z}\right)^{2}}{g^{2}}
  &=&
  \frac{1}{2s}
  +
  \frac{\cos\alpha\left(\cos\alpha + e^{-s}\right)}{(1 + e^{-s}\cos\alpha)^{2}}
  \label{eq:kouchan-A2is1-3.3.12}
  , \\
  g^{2}\left(\Delta z\right)^{2}
  &=&
  \frac{s}{2}
  -
  \frac{s^{2}e^{-s} \cos\alpha}{1 + e^{-s}\cos\alpha}
  \label{eq:kouchan-A2is1-3.3.14}
  .
\end{eqnarray}
We also note that the first term in
Eq.~(\ref{eq:kouchan-A2is1-3.3.12})
[Eq.~(\ref{eq:kouchan-A2is1-3.3.14})] shows the initial 
variance in $p_{z}$ (in $z$).
The remaining terms in Eqs.~(\ref{eq:kouchan-A2is1-3.3.12}) and
(\ref{eq:kouchan-A2is1-3.3.14}) are due to the pre-selection,
weak interaction, and the post-selection.


The probability density of the detector after the post-selection
in $p_{z}$-space is given by  
\begin{eqnarray}
  \langle p_{z}|\rho_{d}'|p_{z}\rangle
  &=&
  \frac{
    \cos\alpha
    + \cosh(2 s p_{z}/g)
    + \sin\alpha \sinh(2 s p_{z}/g)
  }{
    1 + e^{-s} \cos\alpha
  }
  \nonumber\\
  && \quad
  \times
  \exp[-s]
  \sqrt{\frac{s}{\pi g^{2}}}
  \exp\left[
    - s (p_{z}/g)^{2}
  \right]
  .
  \nonumber\\
  \label{eq:kouchan-A2is1-3.3.26-2}
\end{eqnarray}


Lorenzo and Egues~\cite{A.D.Lorenzo-J.C.Egues-2008} also derived
analytical formulae of the expectation of the pointer variable
and the probability distribution in more complicated form.
Their derivation is based on Born's rule of the joint
probability.
Our results shown here are consistent with their results.


\subsection{Expectation value optimization} 
\label{sec:AAV-Expectation-value-optimization}


Figure~\ref{fig:kouchan-AAV-Expect-for-PRA-final}
explicitly shows the existence of the ridge in the surface of
the expectation value (\ref{eq:AAV-expectation-value}).
This means that for a fixed coupling parameter $s$, there
is the optimal pre-selection angle $\alpha$ at which the
expectation value (\ref{eq:AAV-expectation-value}) is maximized.
This was pointed out by Wu and Li~\cite{S.Wu-Y.Li-2010} from the
less accurate expression (\ref{eq:S.Wu-Y.Li-2010-19}).
On the other hand, we can accurately discuss this optimization
of the expectation value from our exact expression
(\ref{eq:AAV-expectation-value}).
Here, we consider this optimization of the expectation value
(\ref{eq:AAV-expectation-value}) in detail.


\begin{figure}[ht]
  \centering
  \includegraphics[width=0.5\textwidth]{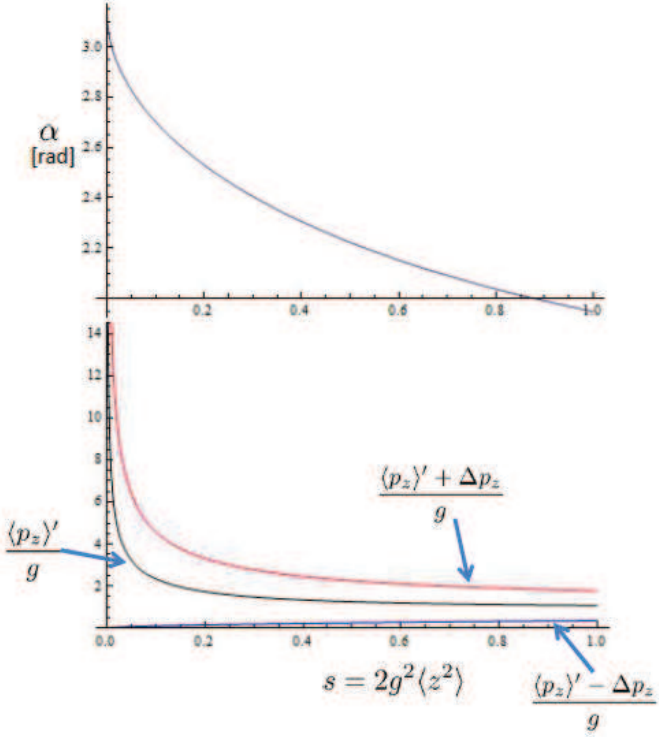}
  \caption{
    {\bf [Online Color]}
    The optimal expectation-value line
    (\ref{eq:AAV-optimal-line}) [top panel], the expectation
    value $\langle p_{z}\rangle'/g$
    [Eq.~(\ref{eq:expectation-pz-on-optimal-line})] and the
    fluctuation $\Delta p_{z}$ 
    [Eq.~(\ref{eq:variance-in-pz-on-optimal-line})] in $p_{z}$
    after the post-selection on the optimal line
    (\ref{eq:AAV-optimal-line}) [lower panel] are shown as
    functions of the coupling parameter $s$.
    We also show that $(\langle p_{z}\rangle'\pm\Delta p_{z})/g$
    in this lower panel.
    [The red line is $(\langle p_{z}\rangle'+\Delta p_{z})/g$ and
    the blue line is $(\langle p_{z}\rangle'-\Delta p_{z})/g$.]
    This figure shows that if we choose the small parameter of
    $s\ll 1$, we can accomplish the large expectation value of
    $\langle p_{z}\rangle'/g\sim 1/\sqrt{2s}$, but the
    fluctuations in $p_{z}$ also amplified as 
    $\Delta p_{z}/g\sim 1/\sqrt{2s}$.
  }
  \label{fig:kouchan-AAV-optimal-PS-PRA-final}
\end{figure}


To derive the points at which the expectation value is
optimized, we consider the equation $\partial(\langle
p_{z}\rangle'/g)/\partial\alpha=0$. 
This equation yields
\begin{eqnarray}
  \label{eq:AAV-optimal-line}
  \cos\alpha = - e^{-s}.
\end{eqnarray}
We call the line which is expressed by
Eq.~(\ref{eq:AAV-optimal-line}) on the $(s,\alpha)$-plane as
the optimal expectation-value line. 
On this optimal line, the expectation value of $p_{z}$ and the
fluctuation $\Delta p_{z}$ in $p_{z}$ are given by 
\begin{eqnarray}
  \label{eq:expectation-pz-on-optimal-line}
  \frac{\langle p_{z}\rangle'}{g}
  &=&
  \frac{1}{\sqrt{1-e^{-2s}}}
  , \\
  \label{eq:variance-in-pz-on-optimal-line}
  \frac{\Delta p_{z}}{g} &=& \frac{1}{\sqrt{2s}}.
\end{eqnarray}
Here, we note that the fluctuation $\Delta p_{z}$
(\ref{eq:variance-in-pz-on-optimal-line}) in $p_{z}$ coincides
with that for the initial state of the detector. 
The optimal expectation-value line, the expectation value
$\langle p_{z}\rangle'$, and the fluctuation $\Delta p_{z}$ in
$p_{z}$ on this optimal line are shown in 
Fig.~\ref{fig:kouchan-AAV-optimal-PS-PRA-final}.


The expectation value (\ref{eq:expectation-pz-on-optimal-line})
of $p_{z}$ explicitly shows that we can accomplish the arbitrary
large weak value amplification if we prepare the sufficiently
small coupling parameter $s$.
Actually, when $s\ll 1$, $\langle p_{z}\rangle'\sim
g/\sqrt{2s}=1/(2\sqrt{\langle z^{2}\rangle})$.
Thus, the expectation value of $p_{z}$ can be very large if we
choose the initial variance in $z$ is very small.
This is just the weak value amplification proposed by
AAV~\cite{Y.Aharonov-D.Z.Albert-L.Vaidman-1988}.


\begin{figure}[ht]
  \centering
  \includegraphics[width=0.5\textwidth]{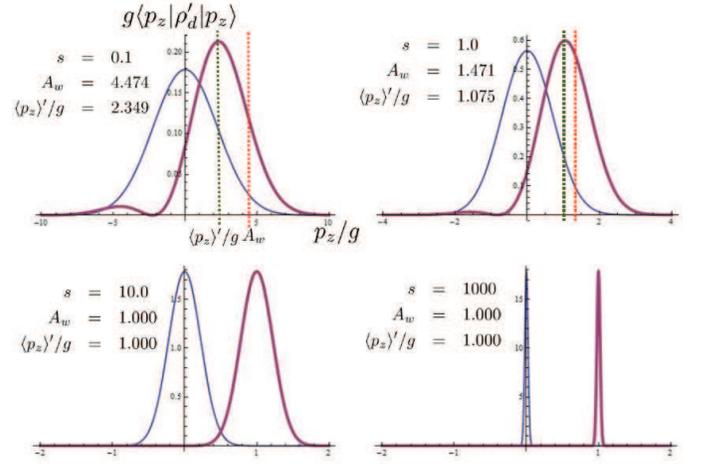}
  \caption{
    {\bf [Online Color]}
    Probability distribution functions
    (\ref{eq:kouchan-AAV-prob-dist-on-OEVL}) in
    $p_{z}/g$-space on the optimal expectation-value line
    (\ref{eq:AAV-optimal-line}) with some coupling parameters
    $s$ are shown by the thick lines (red lines).
    We also plot the initial probability distribution
    $\sqrt{\frac{s}{\pi}} e^{- s (p_{z}/g)^{2}}$ with the
    same coupling parameter $s$ by the thin lines (blue lines).
    $s=0.1$ case corresponds to the weak measurement on the
    optimal expectation-value line (\ref{eq:AAV-optimal-line}).
    This shows that the peak of the probability distribution
    slightly deviates from the weak value $A_{w}$.
    $s=1.0$ case is still essentially same as $s=0.1$ case.
    $s=10$ and $s=1000$ cases correspond to the strong
    measurement case.
    These behaviors of the probability distribution
    (\ref{eq:kouchan-AAV-prob-dist-on-OEVL}) also well-describes 
    the strong measurement regime $s\gg 1$. 
  }
  \label{fig:kouchan-AAV-opt-probdist-PRA-final}
\end{figure}


\begin{figure}
  \centering
  \includegraphics[width=0.5\textwidth]{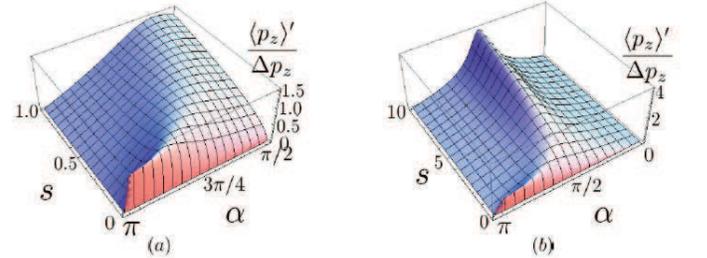}
  \caption{
    The signal to noise ratio (SNR)
    (\ref{eq:kouchan-A2is1-3.3.27}) is shown as a function of
    the coupling parameter $s$ and the pre-selection angle
    $\alpha$ : (a) $s\in(0,1)$ and $\alpha\in(\pi/2,\pi)$; 
    (b) $s\in(0,10)$ and $\alpha\in(\pi,0)$.
    There is the ridge of the SNR from $(s,\alpha)=(0,\pi)$ to
    $(s,\alpha)=(+\infty,\pi/2)$.
    [We note that the direction of $s$-axes are opposite to
    those in Fig.~\ref{fig:kouchan-AAV-Expect-for-PRA-final}.]
    The ridge around $\alpha=\pi/2$ is due to the strong
    measurement regime $s\gg 1$, which is due to the fact that
    $\alpha=\pi/2$ corresponds to the eigenstate
    $|\uparrow_{z}\rangle$ of the operator $\sigma_{z}$ with the
    eigenvalue $+1$.
    Together with
    Fig.~(\ref{fig:kouchan-AAV-opt-probdist-PRA-final}), this
    shows the behavior of the SNR between the weak-measurement
    regime $s\ll 1$ and the strong-measurement regime $s\gg 1$. 
  }
  \label{fig:kouchan-AAV-SNR-for-PRA-final}
\end{figure}


On the optimal expectation-value line
(\ref{eq:AAV-optimal-line}), the probability distribution
(\ref{eq:kouchan-A2is1-3.3.26-2}) is given by 
\begin{eqnarray}
  \langle p_{z}|\rho_{d}'|p_{z}\rangle
  &=&
  \frac{
      \cosh(2 s x)
    + \sqrt{1 - e^{-2s}} \sinh(2 s x)
    - e^{-s}
  }{
    1 - e^{-2s}
  }
  \nonumber\\
  && \quad
  \times
  \sqrt{\frac{s}{\pi g^{2}}}
  e^{- s (1 +x^{2})}
  ,
  \label{eq:kouchan-AAV-prob-dist-on-OEVL}
\end{eqnarray}
where $x:=p_{z}/g$.
This probability distribution on the optimal expectation-value
line is shown in
Fig.~\ref{fig:kouchan-AAV-opt-probdist-PRA-final} with some
coupling parameters $s$.
In Fig.~\ref{fig:kouchan-AAV-opt-probdist-PRA-final}, $s=0.1$
case corresponds to the weak measurement on the optimal
expectation-value line (\ref{eq:AAV-optimal-line}). 
This shows that the peak of the probability distribution
slightly deviates from the weak value $A_{w}$ and the
probability density after the post-selection is slightly
different from the Gaussian
profile~\cite{I.M.Duck-P.M.Stevenson-E.C.G.Sudarshan-1989}. 
$s=1.0$ case is still essentially same as $s=0.1$ case.
$s=10$ and $s=1000$ cases correspond to the strong measurement
regime. 
On the optimal expectation-value line
(\ref{eq:AAV-optimal-line}), $\alpha$ approaches to $\pi/2$ in the
limit $s\rightarrow\infty$ and $\Delta p_{z}$ in
Eq.~(\ref{eq:variance-in-pz-on-optimal-line}) approaches to
$0$.
Here, we note that the pre-selected state with $\alpha=\pi/2$ 
corresponds to the eigenstate $|\uparrow_{z}\rangle$ of
$\sigma_{z}$.
In this case, we measure this eigenvalue $+1$ with small
uncertainty.
This situation is well-described by the behavior of the
probability distribution with $s=1000$ in
Fig.~\ref{fig:kouchan-AAV-opt-probdist-PRA-final}. 
Therefore, the probability distribution
(\ref{eq:kouchan-AAV-prob-dist-on-OEVL}) well-describes not only
in the weak measurement regime $s\ll 1$ but also in the strong
measurement regime $s\gg 1$.


Although we have an arbitrary large expectation value
(\ref{eq:expectation-pz-on-optimal-line}) if we choose $s\ll 1$,
small coupling parameter $s$ gives large variance in $p_{z}$, as
shown in Eq.~(\ref{eq:variance-in-pz-on-optimal-line}). 
Actually, fluctuation $\Delta p_{z}$ in
Eq.~(\ref{eq:variance-in-pz-on-optimal-line}) also behaves as
$\Delta p_{z}=g/\sqrt{2s}=1/(2\sqrt{\langle z^{2}\rangle})$.
Since the fluctuation $\Delta p_{z}$ is regarded as a noise in
this weak measurement, this means that the SNR on the optimal
expectation-value line (\ref{eq:AAV-optimal-line}) is 
$\langle p_{z}\rangle'/\Delta p_{z}\sim 1$.
Thus, we do not have a large SNR in the expectation-value
(signal) optimization of the single particle experiment.  
Therefore, we consider the optimization of the SNR in
the next subsection.


\subsection{SNR optimization} 
\label{sec:AAV-SNR-optimization} 


The expectation value $\langle p_{z}\rangle'$
[Eq.~(\ref{eq:AAV-expectation-value})] and the fluctuation
$\Delta p_{z}:=\sqrt{\langle(p_{z}-\langle
  p_{z}\rangle')^{2}\rangle'}$
[Eq.~(\ref{eq:kouchan-A2is1-3.3.12})] after the post-selection 
are regarded as the signal and a noise in the measurement of
$p_{z}$.
Therefore, in this section, we regard the ratio 
\begin{eqnarray}
  \frac{\langle p_{z}\rangle'}{\Delta p_{z}}
  =
  \frac{
    \sqrt{2s} \sin\alpha
  }{
    \sqrt{
      (1 + e^{-s}\cos\alpha)^{2}
      +
      2s \cos\alpha\left(\cos\alpha + e^{-s}\right)
    }
  }
  \nonumber\\
  \label{eq:kouchan-A2is1-3.3.27}
\end{eqnarray}
as the SNR and we consider the optimization of this SNR.


In Fig.~\ref{fig:kouchan-AAV-SNR-for-PRA-final}, the behavior of
the SNR (\ref{eq:kouchan-A2is1-3.3.27}) is shown as a
function of the coupling parameter $s$ and the pre-selection
angle $\alpha$ in two different ranges of $s$.
We can see that there is the ridge of the SNR from the
weak-measurement regime $(s,\alpha)=(0,\pi)$ to the
strong-measurement regime $(s,\alpha)=(+\infty,\pi/2)$.
In the strong-measurement regime, the fluctuation 
$\Delta p_{z}$ in $p_{z}$ after the post-selection behaves as 
$\Delta p_{z}\sim g/\sqrt{2s}=1/(2\sqrt{\langle z^{2}\rangle})$,
i.e., $\Delta p_{z}$ approach to zero in the limit
$s\rightarrow\infty$, while the signal 
$\langle p_{z}\rangle'\sim g$ in this strong-measurement
regime.
Then, the SNR has the maximum at $\alpha=\pi/2$ in the
strong-measurement regime $s\gg 1$.
As in Fig.~\ref{fig:kouchan-AAV-opt-probdist-PRA-final},
Fig.~\ref{fig:kouchan-AAV-SNR-for-PRA-final} shows the behavior
of the SNR between the weak-measurement regime $s\ll 1$ and the
strong-measurement regime $s\gg 1$.

\begin{figure}[ht]
  \centering
  \includegraphics[width=0.5\textwidth]{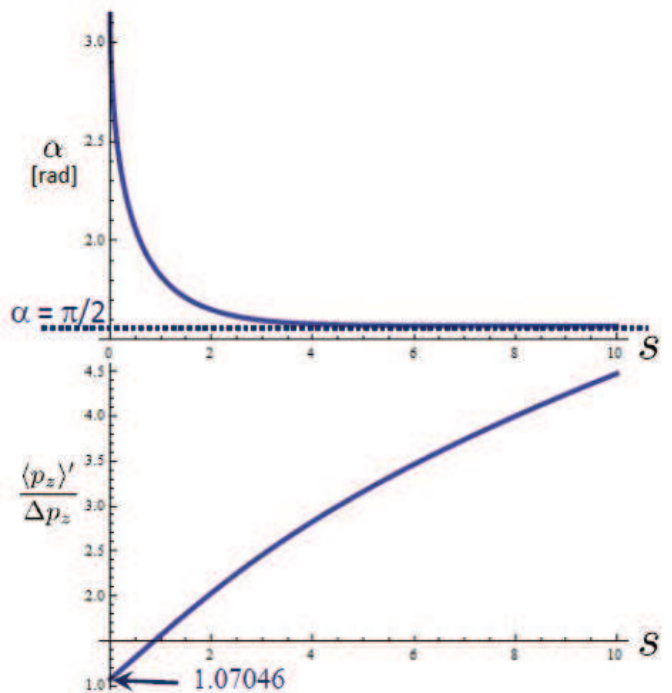}
  \caption{
    The optimal SNR line (upper panel)
    (\ref{eq:kouchan-A2is1-3.3.31}) on the $(s,\alpha)$-plane
    and the optimally pre-selected SNR
    (\ref{eq:kouchan-A2is1-3.3.32}) (lower panel) are shown. 
    The optimally pre-selected SNR is a monotonically increasing 
    function of $s$. 
    In the strong-measurement regime $s\gg 1$, this SNR
    increases due to the fact that pre-selected state
    $\alpha=\pi/2$ is the eigenstate $|\uparrow_{z}\rangle$ of
    the operator $\sigma_{z}$.
    In the weak-measurement regime $s\ll 1$, the SNR cannot be
    larger than that in the strong-measurement regime but has
    the minimum value $\langle p_{z}\rangle'/\Delta 
    p_{z}\sim 1.07046$.
  }
  \label{fig:kouchan-AAV-SNR-optimal-SNR-PRA-final}
\end{figure}

In the both of the weak measurement regime $s\ll 1$ and the
strong measurement regime $s\gg 1$,
Fig.~\ref{fig:kouchan-AAV-SNR-for-PRA-final} implies that, for a
fixed coupling parameter $s$, there is an optimal pre-selection
angle $\alpha$ which maximize the SNR. 
To seek this optimal pre-selection angle, we consider the
equation $\partial(\langle p_{z}\rangle'/\Delta
p_{z})/\partial\alpha=0$.
This equation yields 
\begin{eqnarray}
  \cos^{2}\alpha
  + 2 \frac{\cosh s + s e^{s}}{1+s} \cos\alpha
  + 1
  = 0
  .
  \label{eq:kouchan-A2is1-3.3.29}
\end{eqnarray}
Taking care of $\cos\alpha\leq 1$, we easily see that the
solution to the optimal SNR equation
(\ref{eq:kouchan-A2is1-3.3.29}) is given by
\begin{eqnarray}
  \cos\alpha 
  &=&
  - \frac{\cosh s + s e^{s}}{s + 1}
  +
  \sqrt{
    \left(\frac{\cosh s + s e^{s}}{s + 1}\right)^{2}
    - 1
  }
  \nonumber\\
  &=:&
  \cos\alpha_{\mbox{opt}}(s)
  \label{eq:kouchan-A2is1-3.3.31}
  .
\end{eqnarray}
This solution $\alpha_{\mbox{opt}}(s)$ is the pre-selection
angle $\alpha$ which optimizes the SNR
(\ref{eq:kouchan-A2is1-3.3.27}) and represents the line on the 
$(s,\alpha)$-plane. 
We call this line as the optimal-SNR line. 
On this line, we can evaluate the optimally pre-selected SNR as
\begin{eqnarray}
  \left.\frac{\langle p_{z}\rangle'}{\Delta p_{z}}(s,\alpha)\right|_{\mbox{opt}}
  &=&
  \frac{\langle p_{z}\rangle'}{\Delta p_{z}}(s,\alpha_{\mbox{opt}})
  .
  \label{eq:kouchan-A2is1-3.3.32}
\end{eqnarray}
The optimal SNR line (\ref{eq:kouchan-A2is1-3.3.31}) on
$(s,\alpha)$-plane and the optimally pre-selected SNR is
shown in
Fig.~\ref{fig:kouchan-AAV-SNR-optimal-SNR-PRA-final}.


As shown in
Fig.~\ref{fig:kouchan-AAV-SNR-optimal-SNR-PRA-final}, in the
strong-measurement regime $s\gg 1$, this SNR increases due to
the fact that pre-selected state $\alpha=\pi/2$ is the
eigenstate $|\uparrow_{z}\rangle$ of the operator $\sigma_{z}$.
On the other hand, in the weak-measurement regime $s\ll 1$, the
SNR cannot be larger than that in the strong-measurement regime
but has the minimum value on the optimal SNR line. 
Actually, for $s\ll 1$, the asymptotic expansion of
Eq.~(\ref{eq:kouchan-A2is1-3.3.32}) yields
\begin{eqnarray}
  \frac{\langle p_{z}\rangle'}{\Delta p_{z}}
  &=&
  \sqrt{\frac{2}{\sqrt{3}}}
  +
  \frac{1}{3} \sqrt{1 + \frac{2}{\sqrt{3}}} s
  +
  O(s^{2})
  ,
  \label{eq:kouchan-A2is1-3.3.33}
\end{eqnarray}
which is larger than $\sqrt{2/\sqrt{3}}\sim 1.0746$.


\section{Application to the simplified DSJH setup} 
\label{sec:Application_to_DSJH_setup}


In this section, we apply our formulae, which are summarized in 
Sec.~\ref{sec:A2is1case-non-orthogonal}, to the simplified 
DSJH~\cite{P.B.Dixon-D.J.Starling-A.N.Jordan-J.C.Howell-2009} 
setup.
We discuss the optimization of the expectation value
of transverse deflections of an optical beam in
Sec.~\ref{sec:sDSJH-Expectation-value-optimization} 
and the optimization of the SNR in
Sec.~\ref{sec:sDSJH-SNR-optimization}.


\subsection{Simplified setup of experiment}
\label{sec:sDSJH_setup_typical}


The simplified version of the DSJH experiment is the measurement
of the tiny tilt of the piezo-driven mirror in a Sagnac
interferometer~\cite{P.B.Dixon-D.J.Starling-A.N.Jordan-J.C.Howell-2009}. 
In
Ref.~\cite{P.B.Dixon-D.J.Starling-A.N.Jordan-J.C.Howell-2009},
they use the which-path information of a photon in the Sagnac
interferometer, which is represented by the photon states
$|\circlearrowright\rangle$ and $|\circlearrowleft\rangle$. 
Here, $|\circlearrowright\rangle$ ($|\circlearrowleft\rangle$)
is the state of a photon which propagates along the clockwise
(counter-clockwise) direction in the Sagnac interferometer.
As the pre-selected state $|\psi_{i}\rangle$ of a photon, they
choose 
\begin{eqnarray}
  \label{eq:sDSJH-pre-selection}
  |\psi_{i}\rangle = \frac{1}{\sqrt{2}} \left(
    i e^{i\phi/2}|\circlearrowleft\rangle
    +
    e^{-i\phi/2}|\circlearrowright\rangle
  \right)
  ,
\end{eqnarray}
where $\phi$ is the phase difference of the states
$|\circlearrowright\rangle$ and $|\circlearrowleft\rangle$
introduced by a Soleil-Babinet compensator.


The weak interaction in the weak measurement is described by the
interaction Hamiltonian  
\begin{eqnarray}
  \label{eq:sDSJH-von-Neumann-interaction}
  {\cal H} = k x\otimes{\bf A}\delta(t-t_{0}),
\end{eqnarray}
where $k$ is the momentum shift of the light path by the tilt of 
the piezo-driven mirror and $x$ represents the shift of the
light image at the dark port of the interferometer.
The quantum operator ${\bf A}$ in
Eq.~(\ref{eq:sDSJH-von-Neumann-interaction}) is given by 
\begin{eqnarray}
  \label{eq:which-path-operator-in-sDSJH}
  {\bf A}
  =
  |\circlearrowright\rangle\langle\circlearrowright|
  -
  |\circlearrowleft\rangle\langle\circlearrowleft|
  .
\end{eqnarray}
We note that the operator ${\bf A}$ satisfy the property 
${\bf A}^{2}=1$.
Then, we may apply our formulae given in
Sec.~\ref{sec:All-order_evaluation_of_WM_for_A2is1}.


As the post-selection of a photon state, we choose the dark-port
in the Sagnac interferometer
\begin{eqnarray}
  \label{eq:sDSJH-post-selection}
  |\psi_{f}\rangle = \frac{1}{\sqrt{2}} \left(
    |\circlearrowleft\rangle
    +
    i |\circlearrowright\rangle
  \right)
  ,
\end{eqnarray}
and the weak value in this setup is given by 
\begin{eqnarray}
  \label{eq:sDSJH-weak-value}
  \langle{\bf A}\rangle_{w}
  =
  \frac{
    \langle\psi_{f}|{\bf A}|\psi_{i}\rangle
  }{
    \langle\psi_{f}|\psi_{i}\rangle
  }
  =
  - i \cot\frac{\phi}{2}
  .
\end{eqnarray}
where $\phi$ is the pre-selection angle in
Eq.~(\ref{eq:sDSJH-pre-selection}). 
We note that this weak value (\ref{eq:sDSJH-weak-value}) is pure 
imaginary.


Comparing Eq.~(\ref{eq:sDSJH-von-Neumann-interaction}) with
Eq.~(\ref{eq:von-Neumann-interaction}), we find the
correspondence of variables as
\begin{eqnarray}
  \label{eq:sDSJH-correspondence}
  p \rightarrow x, \quad
  q \rightarrow - p, \quad
  g \rightarrow k,
\end{eqnarray}
where new variables $x$ and $p$ satisfy the commutation relation
$[x,p]=i$.


Although Dixon et al. modified the beam radius of the laser by
lenses in
Ref.~\cite{P.B.Dixon-D.J.Starling-A.N.Jordan-J.C.Howell-2009},
we do not take account of the effect of this modification.
This modification is not essential to the basic mechanism of the
weak measurement. 
This is the reason why we call the ``simplified'' DSJH setup in
this paper.


\subsection{All-order expectation values and probability distribution} 
\label{sec:sDSJH-All-order} 


Here, we apply the formulae summarized in
Sec.~\ref{sec:A2is1case-non-orthogonal} to the above simplified
DSJH setup.


The normalization factor ${\cal Z}$
[Eq.~(\ref{eq:kouchan-A2is1-10-2})], is given by  
\begin{eqnarray}
  {\cal Z}
  &=&
  \frac{1 - e^{-s} \cos\phi}{1-\cos\phi}
  ,
  \label{eq:Dixon-et-al-2009-kouchan-19}
\end{eqnarray}
where $s$ is the coupling parameter [see
Eq.~(\ref{eq:coupling-parameter-def})] defined by 
\begin{eqnarray}
  s := 2 k^{2}\langle x^{2}\rangle.
  \label{eq:Dixon-et-al-2009-kouchan-s-def}
\end{eqnarray}
The expectation values of $x$ and $p$ after the post-selection 
are given by
\begin{eqnarray}
  \label{eq:Dixon-et-al-2009-kouchan-20}
  k \langle x\rangle'
  =
  -
  \frac{
    s e^{-s} \sin\phi
  }{
    1
    - e^{-s} \cos\phi
  }
  ,
  \quad
  \langle p\rangle'
  =
  0
  .
\end{eqnarray}
Fluctuations 
$\Delta x:=\sqrt{\langle (x - \langle x\rangle')^{2}\rangle'}$
and 
$\Delta p:=\sqrt{\langle (p - \langle p\rangle')^{2}\rangle'}$
in $x$ and $p$ are given by
\begin{eqnarray}
  k^{2}(\Delta x)^{2}
  &=&
  \frac{s}{2}
  +
  \frac{
    s^{2}e^{-s}
    \left(
      \cos\phi - e^{-s}
    \right)
  }{
    \left(1 - e^{-s} \cos\phi\right)^{2}
  }
  ,
  \label{eq:Dixon-et-al-2009-kouchan-22}
  \\
  \frac{1}{k^{2}} (\Delta p)^{2}
  &=&
  \frac{1}{2s}
  + \frac{
    1
  }{
    1 - e^{-s} \cos\phi
  }
  .
  \label{eq:Dixon-et-al-2009-kouchan-23}
\end{eqnarray}
As in the case of AAV setup, the first term in
Eq.~(\ref{eq:Dixon-et-al-2009-kouchan-22})
[Eq.~(\ref{eq:Dixon-et-al-2009-kouchan-23})] shows the initial 
variance in $x$ (in $p$).
The remaining terms in
Eqs.~(\ref{eq:Dixon-et-al-2009-kouchan-22}) and
(\ref{eq:Dixon-et-al-2009-kouchan-23}) are due to the
pre-selection, weak interaction, and the post-selection.

\begin{figure}[ht]
  \centering
  \includegraphics[width=0.5\textwidth]{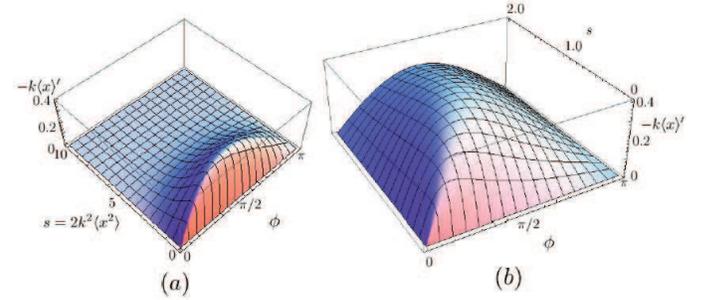}
  \caption{
    $-k\langle x\rangle'$
    [Eq.~(\ref{eq:Dixon-et-al-2009-kouchan-20})] is shown as a
    function of the coupling $s=2k^{2}\langle x^{2}\rangle$ and
    the pre-selection angle $\phi$ in two ranges of $s$. 
    (a) $s\in(0,10)$, (b) $s\in(0,2)$.
    For large $s$, the expectation value $-k\langle x\rangle'$
    decays exponentially.
    From these figures, we can see that, for a given coupling 
    parameter $s$, there is a pre-selection angle $\phi$ such
    that the expectation value of $x$ is maximized.
  }
  \label{fig:kouchan-Dixon-x-expect-for-PRA-final}
\end{figure}

In Fig.~\ref{fig:kouchan-Dixon-x-expect-for-PRA-final}, the
expectation value $-k\langle x\rangle'$ of
Eq.~(\ref{eq:Dixon-et-al-2009-kouchan-20}) is shown as a 
function of the coupling parameter 
$s=2k^{2}\langle x^{2}\rangle'$ and the pre-selection angle
$\phi$.


In the simplified DSJH setup, the probability density in
$x$-space is obtained from Eq.~(\ref{eq:kouchan-A2is1-101}) as
\begin{eqnarray}
  \langle x|\rho_{d}'|x\rangle
  &=&
  \frac{
    1
    - \cos(2kx-\phi)
  }{
    1
    - e^{-s} \cos\phi
  }
  \langle x|\rho_{d}|x\rangle
  ,
  \label{eq:Dixon-et-al-2009-kouchan-51}
\end{eqnarray}
where $\langle x|\rho_{d}|x\rangle$ is the initial probability
density in $x$-space:
\begin{eqnarray}
  \langle x|\rho_{d}|x\rangle
  =
  \frac{k}{\sqrt{\pi s}}
  \exp\left[- \frac{k^{2}x^{2}}{s}\right]
  \label{eq:Dixon-et-al-2009-kouchan-initial-prob-density-x}
  .
\end{eqnarray}


\subsection{Expectation value optimization}
\label{sec:sDSJH-Expectation-value-optimization}

\begin{figure}[ht]
  \centering
  \includegraphics[width=0.5\textwidth]{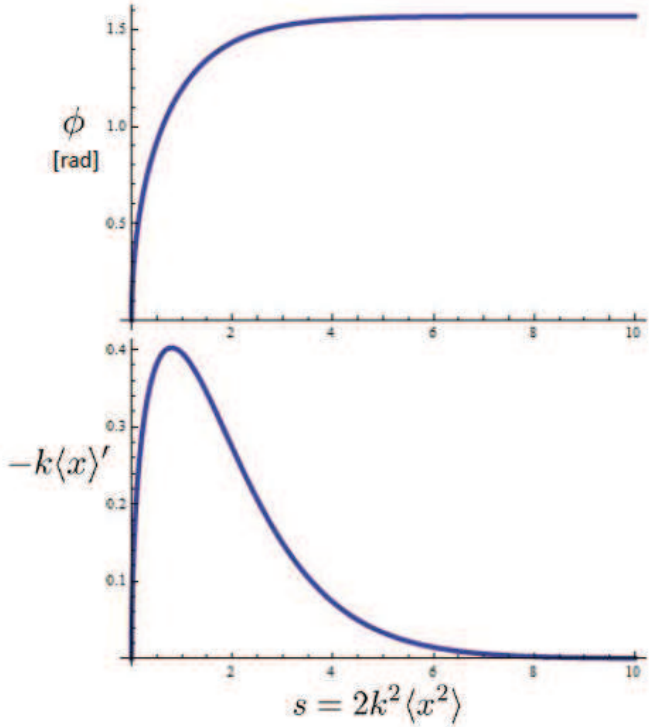}
  \caption{
    The optimal expectation-value line
    (\ref{eq:Dixon-et-al-2009-kouchan-29}) (upper panel) and the  
    expectation value of $-k\langle x\rangle'$ 
    (\ref{eq:Dixon-et-al-2009-kouchan-30-optimal-exp}) (lower
    panel) on this optimal line is shown as a function of the
    coupling parameter $s$.
    In the limit $s\rightarrow\infty$, the optimal pre-selection
    angle for expectation value approaches to
    $\phi\rightarrow\pi/2$. 
    Further, on the optimal expectation-value line, the
    expectation value $-k\langle x\rangle'$ has the maximum
    value at $s\sim 0.8$.
  }
  \label{fig:kouchan-Dixon-optimal-line-final}
\end{figure}

Here, we consider the optimization of the expectation value
amplification in the simplified DSJH setup.
From Fig.~\ref{fig:kouchan-Dixon-x-expect-for-PRA-final}, we can
see that the expectation value 
(\ref{eq:Dixon-et-al-2009-kouchan-20}) exponentially decays in
the strong measurement regime $s\gg 1$
(Fig.~\ref{fig:kouchan-Dixon-x-expect-for-PRA-final}(a)).
Furthermore,
Fig.~\ref{fig:kouchan-Dixon-x-expect-for-PRA-final}(b) also
shows that, for a given coupling parameter $s$, there is a
pre-selection angle $\phi$ such that the expectation value of $x$
is maximized.
This is the optimal expectation value of $-k\langle x\rangle'$
for a fixed coupling parameter $s$.
To seek this optimal expectation value, we consider the equation
$\partial(-k\langle x\rangle')/\partial\phi=0$, which yields the
equation 
\begin{eqnarray}
  \cos\phi = e^{-s}.
  \label{eq:Dixon-et-al-2009-kouchan-29}
\end{eqnarray}
This is the equation for the optimal expectation-value line on
$(\phi,s)$-plane. 
On this optimal line, the expectation-value of $x$ and the
fluctuation $\Delta x$ in $x$ are given by 
\begin{eqnarray}
  k \langle x\rangle'
  &=&
  -
  \frac{
    s e^{-s}
  }{
    \sqrt{1 - e^{-2s}}
  }
  =:
  k \langle x\rangle'_{\mbox{opt}}
  ,
  \label{eq:Dixon-et-al-2009-kouchan-30-optimal-exp}
  \\
  k \Delta x
  &=&
  \sqrt{\frac{s}{2}}
  .
  \label{eq:Dixon-et-al-2009-kouchan-22-optimal-var}
\end{eqnarray}
We note that the variance
[Eq.~(\ref{eq:Dixon-et-al-2009-kouchan-22-optimal-var})] in $x$
after the optimal post-selection coincides with that of the
initial state of the detector.
The optimal expectation-value line
(\ref{eq:Dixon-et-al-2009-kouchan-29}) and the expectation value
of $-k\langle x\rangle'$
(\ref{eq:Dixon-et-al-2009-kouchan-30-optimal-exp}) on this
optimal line is shown in
Fig.~\ref{fig:kouchan-Dixon-optimal-line-final}.


\begin{figure}[ht]
  \centering
  \includegraphics[width=0.5\textwidth]{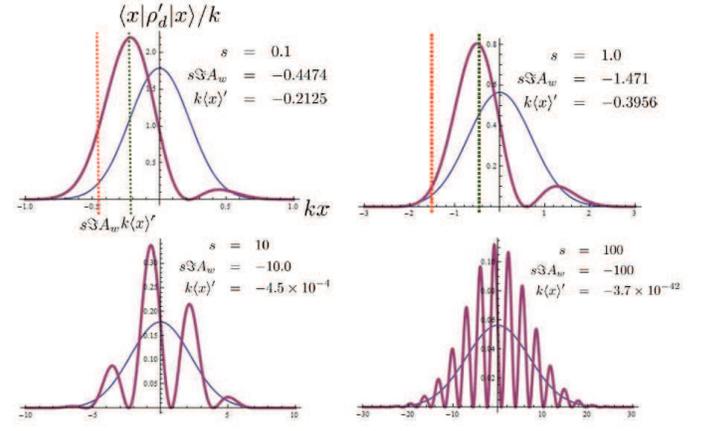}
  \caption{
    {\bf [Online Color]}
    Probability distribution functions
    (\ref{eq:Dixon-et-al-2009-kouchan-51-optimal-line}) in
    $x$-space on the optimal expectation-value line
    (\ref{eq:Dixon-et-al-2009-kouchan-29}) with some coupling 
    parameters $s$ are shown by the thick line (red line).
    We also plot the initial probability distribution
    $\langle x|\rho_{d}\rangle/k$ in
    Eq.~(\ref{eq:Dixon-et-al-2009-kouchan-initial-prob-density-x})
    with the same coupling parameter $s$ by the thin lines (blue
    lines).
    The $s=0.1$ case corresponds to the weak measurement on the 
    optimal expectation-value line
    (\ref{eq:Dixon-et-al-2009-kouchan-29}). 
    This shows that the peak of the probability distribution
    slightly deviates from the linear result $s\Im A_{w}$.
    The $s=1.0$ case is still essentially same as $s=0.1$ case.
    The maximal expectation value ($s=s_{m}\sim 0.8$) is
    obtained around this parameter.
    The $s=10$ and $s=1000$ cases correspond to the strong
    measurement case.
  }
  \label{fig:kouchan-Dixon-opt-probdist-PRA-final}
\end{figure}

From Eq.~(\ref{eq:Dixon-et-al-2009-kouchan-51}), the probability
density on the optimal expectation-value line
(\ref{eq:Dixon-et-al-2009-kouchan-29}) is given by 
\begin{eqnarray}
  \langle x|\rho_{d}'|x\rangle
  &=&
  \frac{
    1
    - e^{-s} \cos(2kx)
    - \sqrt{1 - e^{-2s}} \sin(2kx)
  }{
    1 - e^{-2s}
  }
  \nonumber\\
  && \quad\quad\quad
  \times
  \frac{k}{\sqrt{\pi s}}
  \exp\left[- \frac{k^{2}x^{2}}{s}\right]
  ,
  \label{eq:Dixon-et-al-2009-kouchan-51-optimal-line}
\end{eqnarray}
which is shown in
Fig.~\ref{fig:kouchan-Dixon-opt-probdist-PRA-final}.
The $s=0.1$ case, which corresponds to the weak measurement on
the optimal expectation-value line
(\ref{eq:Dixon-et-al-2009-kouchan-29}), shows that the peak of
the probability distribution 
slightly deviates from the linear result 
$s\Im A_{w}$ given in
Ref.~\cite{P.B.Dixon-D.J.Starling-A.N.Jordan-J.C.Howell-2009},
and that the probability density after the post-selection is
slightly different from the Gaussian distribution.
When the coupling parameter $s$ is large, many peaks appear in
the probability density in $x$-space and the expectation 
value $\langle x\rangle'$ after the post-selection approaches to
zero due to the contribution of these many peaks.


In the limit $s\rightarrow\infty$, the optimal expectation-value
line approaches to $\phi\rightarrow\pi/2$.
Further, we have to note that, on the optimal expectation-value 
line, the expectation value $-k\langle x\rangle'$ has the
maximum value at $s\sim 0.8$.
This is maximal value of $-k\langle x\rangle'$ on whole
$(\phi,s)$-plane.
To seek this maximum point, we consider the equation
$\partial(k \langle x\rangle'_{\mbox{opt}})/\partial s = 0$.
The solution $s=s_{m}$ to this equation is derived from the
equation 
\begin{eqnarray}
    1 - s_{m} - e^{-2s_{m}} = 0
    .
  \label{eq:Dixon-et-al-2009-kouchan-36}
\end{eqnarray}
The numerical value of $s_{m}$ is $s_{m} \simeq 0.79681$.
Therefore, the expectation value satisfy the inequality 
\begin{eqnarray}
  - k\langle x\rangle'
  \leq
  \left.- k \langle x\rangle'_{opt}\right|_{s=s_{m}}
  \simeq 0.402371.
  \label{eq:Dixon-et-al-2009-kouchan-maixmal-exp-value}
\end{eqnarray}
At this maximum point, the optimal pre-selection angle
$\phi_{m}$ is determined by $\cos\phi_{m}=e^{-s_{m}}$, which
yields $\phi_{m}\simeq 1.103$ rad $\simeq 63.2^{\circ}$.
We also note that the expectation value $\langle x\rangle'$
at the maximum point $s=s_{m}$ itself is proportional to
$k^{-1}$. 
Therefore, we can obtain the large expectation value 
$\langle x\rangle'$ if we have a small coupling constant $k$ in
the interaction Hamiltonian
(\ref{eq:sDSJH-von-Neumann-interaction}).


If we evaluate the amplification factor ${\cal A}$ by following
to the discussion by Dixon et 
al.~\cite{P.B.Dixon-D.J.Starling-A.N.Jordan-J.C.Howell-2009}, 
the amplification factor is given by 
${\cal A}=|\langle x\rangle'|/\delta$.
Here, $\delta=kl_{md}/k_{0}$ is the unamplified deflection
without the interferometer.
The unamplified deflection in their experiment is $\delta\sim 3$
$\mu$m.
On the other hand, from
Eq.~(\ref{eq:Dixon-et-al-2009-kouchan-maixmal-exp-value}), we
obtain $|\langle x\rangle'_{opt}|_{s=s_{m}}|\simeq 0.4/k$ at the
maximum point $s=s_{m}$.
Since $k\sim 2 \times 10^{-5}$ $\mu$m$^{-1}$ in their experiment,
the maximal amplification is estimated as ${\cal A}_{max} \sim 600$.


However, since the definition of $s$ is given by
Eq.~(\ref{eq:Dixon-et-al-2009-kouchan-s-def}) and the
fluctuation $\sqrt{\langle x^{2}\rangle}$ of the initial
variance is regarded as the beam radius, $s_{m}\sim 0.8$
corresponds to $\sqrt{\langle x^{2}\rangle} \sim 0.63 \times
(1/k)$.
The optimal beam radius in their setup is given by
$\sqrt{\langle x^{2}\rangle} \sim 3$ cm from 
Eq.~(\ref{eq:Dixon-et-al-2009-kouchan-maixmal-exp-value}).
On the other hand, the maximum expectation value 
$-\langle x\rangle'\sim 0.4\times(1/k)\sim 2$ cm.
For one-photon case, the SNR at the maximal point $s=s_{m}$ is
given by $|\langle x\rangle'|/\Delta x\sim 0.6$, which is
independent of the coupling constant $k$ in the interaction
Hamiltonian. 
For this reason, in Sec.~\ref{sec:sDSJH-SNR-optimization}, we
consider the optimization of this SNR.


\subsection{SNR optimization}
\label{sec:sDSJH-SNR-optimization}

\begin{figure}
  \centering
  \includegraphics[width=0.5\textwidth]{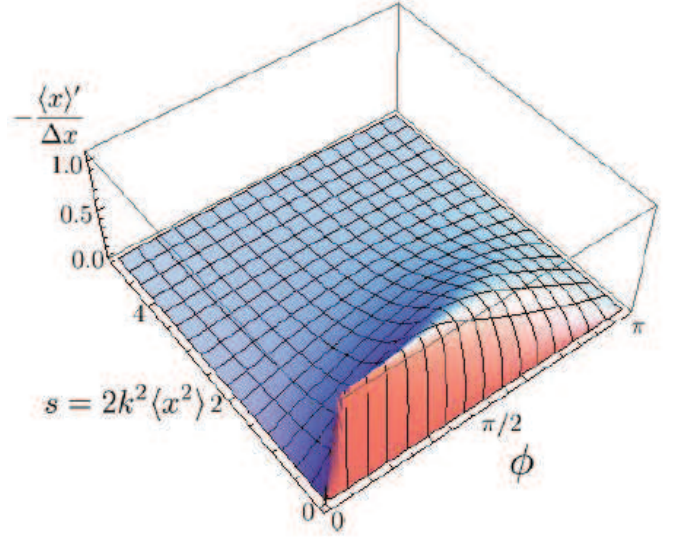}
  \caption{
    The signal to noise ratio (SNR)
    (\ref{eq:Dixon-et-al-2009-kouchan-52-1}) in simplified DSJH
    setup are shown as a function of the coupling parameter
    $s$ and the pre-selection angle $\phi$.
    This shows that the SNR in this setup have the peak only in
    the weak measurement regime $s<1$.
  }
  \label{fig:kouchan-Dixon-x-SNR-surface-PRA-final}
\end{figure}

As in the case of the AAV setup, we consider the optimization of
the SNR.
From Eqs.~(\ref{eq:Dixon-et-al-2009-kouchan-20}) and
(\ref{eq:Dixon-et-al-2009-kouchan-22}), the SNR is given by 
\begin{eqnarray}
  \frac{\left|\langle x\rangle'\right|}{\Delta x}
  =
  \frac{
    \sqrt{2s} e^{-s} \sin\phi
  }{
    \sqrt{
      \left(1 - e^{-s} \cos\phi\right)^{2}
      +
      2 s e^{-s} \left( \cos\phi - e^{-s} \right)
    }
  }
  .
  \nonumber\\
  \label{eq:Dixon-et-al-2009-kouchan-52-1}
\end{eqnarray}
The behavior of this SNR on $(\phi,s)$-plane is shown in
Fig.~\ref{fig:kouchan-Dixon-x-SNR-surface-PRA-final}, which
indicates that the SNR (\ref{eq:Dixon-et-al-2009-kouchan-52-1})
in the simplified DSJH setup is maximized only in the weak
measurement regime $s<1$.


\begin{figure}
  \centering
  \includegraphics[width=0.5\textwidth]{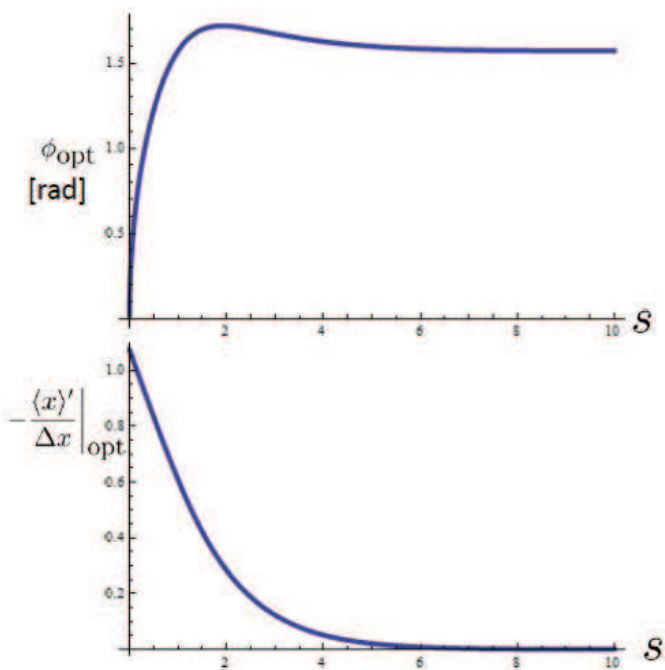}
  \caption{
    The optimal-SNR pre-selection angle $\phi_{\mbox{opt}}$
    [the solution to
    Eq.~(\ref{eq:Dixon-et-al-2009-kouchan-62-a})] (upper panel)
    and the optimized SNR
    (\ref{eq:Dixon-2009-kouchan-optimal-SNR}) (lower panel) in
    simplified DSJH setup are shown as a function of the
    coupling parameter $s$.
    This shows that the SNR in this setup have the peak only in
    the weak measurement regime $s<1$.
    The maximum SNR is of the order of unity in the single
    photon case. 
  }
  \label{fig:kouchan-Dixon-x-optimal-SNR}
\end{figure}

To carry out the optimization of the SNR
(\ref{eq:Dixon-et-al-2009-kouchan-52-1}), we consider the
equation 
$\partial(|\langle x\rangle'|/\Delta x)/\partial\phi=0$, which  
yields
\begin{eqnarray}
  \cos^{2}\phi
  + 2 \frac{(\cosh s - s e^{-s})}{s - 1}\cos\phi
  + 1
  =
  0
  .
  \label{eq:Dixon-et-al-2009-kouchan-54-1}
\end{eqnarray}
The solution $\phi=\phi_{\mbox{opt}}$ to
Eq.~(\ref{eq:Dixon-et-al-2009-kouchan-54-1}) is given by
\begin{eqnarray}
  \cos\phi_{\mbox{opt}}(s)
  &:=&
  - \frac{\cosh s - s e^{-s}}{s - 1}
  \nonumber\\
  &&
  +
  \frac{
    \sqrt{
      (\cosh s - s e^{-s})^{2} - (s - 1)^{2}
    }
  }{
    s - 1
  }
  ,
  \nonumber\\
  \label{eq:Dixon-et-al-2009-kouchan-62-a}
\end{eqnarray}
Equation (\ref{eq:Dixon-et-al-2009-kouchan-62-a}) describes the
optimal line on $(\phi,s)$-plane.  
On this optimal line, the optimal SNR is given by 
\begin{eqnarray}
  \left.\frac{\left|\langle x\rangle'\right|}{\Delta x}\right|_{\mbox{opt}}
  := 
  \left.\frac{\left|\langle x\rangle'\right|}{\Delta x}\right|_{\phi=\phi_{\mbox{opt}}}.
  \label{eq:Dixon-2009-kouchan-optimal-SNR}
\end{eqnarray}
Along the optimal line
[Eq.~(\ref{eq:Dixon-et-al-2009-kouchan-62-a})], the optimized
SNR is shown in Fig.~\ref{fig:kouchan-Dixon-x-optimal-SNR}. 
The optimal SNR (\ref{eq:Dixon-2009-kouchan-optimal-SNR}) is a
monotonically decreasing function of $s$.
Further, only in the region $s<0.15$, the optimal SNR
(\ref{eq:Dixon-2009-kouchan-optimal-SNR}) can be larger than
unity. 
Actually, for $s\ll 1$, the asymptotic behavior of the optimal
SNR (\ref{eq:Dixon-2009-kouchan-optimal-SNR}) is given by 
\begin{eqnarray}
  \left.\frac{\left|\langle x\rangle'\right|}{\Delta x}\right|_{\mbox{opt}}
  &=&
  \sqrt{\frac{2}{\sqrt{3}}}
  -
  \frac{1}{3} \sqrt{1 + \frac{2}{\sqrt{3}}} s 
  +
  O(s^{2})
  \nonumber\\
  &<&
  \sqrt{\frac{2}{\sqrt{3}}},
  \label{eq:Dixon-2009-kouchan-optimal-SNR-for-small-s}
\end{eqnarray}
where $\sqrt{2/\sqrt{3}}\simeq 1.0746$.
Thus, we have shown that the upper limit of the SNR in the
simplified DSJH setup for the
single photon case is of the order of unity.


\section{Summary}
\label{sec:Summary_Discussion}


In summary, after reviewing the formulation by Wu and
Li~\cite{S.Wu-Y.Li-2010}, we derived some formulae for the
weak measurement of the operator ${\bf A}$ which satisfies 
the property ${\bf A}^{2}=1$ through their formulation.
We have to emphasize that our formulae are based on the
exact evaluation of the formulation of Wu and Li.
In the derivation of these formulae, we assume that the initial
state of the detector is zero mean-value Gaussian.
We note that we do not use any additional condition to derive
these formulae.
Our formulae are valid not only in the weak measurement regime
but also in the strong measurement regime. 
Due to this fact, we could clarify the connection between the
strong measurement regime and the weak measurement regime.


We applied our formulae to two experimental setups.
One is the experiment of the weak measurement using spin-1/2
particles, which was proposed in AAV original paper of the weak 
measurement~\cite{Y.Aharonov-D.Z.Albert-L.Vaidman-1988}.
The other is the simplified version of the optical experiment
in the Sagnac interferometer (simplified DSJH setup) by Dixon 
et al.~\cite{P.B.Dixon-D.J.Starling-A.N.Jordan-J.C.Howell-2009}.
These two experimental setups are typical experiments of the
weak measurements.
The weak value is real in the AAV setup, while it is pure
imaginary in the simplified DSJH setup.
In these setups, we have two control parameters.
One is the pre-selection angles in these experiment and the
other is the coupling parameter $s$ defined by
Eq.~(\ref{eq:coupling-parameter-def}).
We discussed the behavior of the expectation values of the
detector variables in the whole range of these two parameters.


In both setups of AAV and DSJH, we found that for a fixed 
coupling parameter $s$, there exits the pre-selection (or
post-selection) which maximize the expectation values of
variables for the detector or the SNR.
The precise estimation of this optimal pre-selection (or
post-selection) is possible through the exact expression of the 
expectation values summarized in this paper.
This is the main results of this paper.
The existence of this optimal pre-selection (or post-selection)
comes from the fact that we specify the subensemble of system
through the post-selection in weak measurements.
Since the post-selection is the restriction of the system
ensemble, the density matrix of the detector after the
postselection is renormalized by this restriction.
This is the essential reason of the appearance of the
normalization factor ${\cal Z}$ in
Eq.~(\ref{eq:total-trace-of-PS-density-matrix-Z}).
The behavior of the normalization factor ${\cal Z}$ leads the
existence of this optimal pre-selection (or post-selection).


Furthermore, we showed that the optimized SNR is order of unity
in the weak measurement regime for the single particle (or
photon) experiment in both experimental setups.
To improve this SNR, we have to consider the large $N$ ensemble
of particles (or photons).
Due to this large $N$ ensemble, the SNR is improved by the
factor $\sqrt{N}$ as proposed by Aharonov and
Vaidman~\cite{Aharonov-Vaidman-2007-update-review}.
In particular, the photon number is very large in the
experiments using the laser beam (for example, the experiment by
Dixon et 
al.~\cite{P.B.Dixon-D.J.Starling-A.N.Jordan-J.C.Howell-2009}).  
For this reason, the large SNR should be obtained in the actual
experiments.


Finally, we have to emphasize that many other experiments are 
also categorized into the case of ${\bf A}^{2}=1$ and the initial
Gaussian state of the detector.
For example, the experiment by Iinuma et
al.~\cite{M.Iinuma-Y.Suzuki-G.Taguchi-Y.Kadoya-H.F.Hofmann-2011}
corresponds to the experiment to measure the operator ${\bf A}$,
which satisfies ${\bf A}^{2}=1$, with a real weak value.
They experimentally confirmed the formula
(\ref{eq:kouchan-A2is1-52}). 
The experiment by Hosten and Kwiat~\cite{O.Hosten-P.Kwiat-2008}
corresponds to the experiment of the operator ${\bf A}$, which 
satisfies ${\bf A}^{2}=1$, with a weak value of pure imaginary. 
Thus, we may say that there are many experiments to which our
formulae are applicable.
Of course, in some actual experiments, there are some
complexity which we did not take into account in this paper.
For example, the modification of the beam radius by lenses in
DSJH
experiment~\cite{P.B.Dixon-D.J.Starling-A.N.Jordan-J.C.Howell-2009}
was not included in our treatment. 
Furthermore, we might have to care about the validity of the
von-Neumann interaction model
(\ref{eq:AAV-von-Neumann-interaction}) in real experiments. 
Although there are some issues to be taken into account when we 
apply our arguments to specific experiments, we expect that our
exact expressions of some expectation values in a weak
measurement will be useful to understand experimental results
or to propose some new experimental setups.


\section*{Acknowledgments}


The authors would like to thanks to all participants of the QND
seminar at National Astronomical Observatory of Japan for
valuable discussions.
A.~N.~ is supported by a Grant-in-Aid through JSPS.


\appendix


\section{Derivations of formulae} 
\label{sec:Derivations_of_formulae}


In this appendix, we show the derivations of the formulae
summarized in
Sec.~\ref{sec:All-order_evaluation_of_WM_for_A2is1}. 
Our derivation use the Wu-Li formulation~\cite{S.Wu-Y.Li-2010}
reviewed in Sec.~\ref{sec:Wu-Li_Formalism}.
Since their formulation requires the separate treatments
according to the norm of the pre- and post-selected states, we
first consider, in
Sec.~\ref{sec:Derivations_of_formulae_non_orthogonal}, the
non-orthogonal case in which the pre- and post-selection is not
orthogonal. 
Then, in Sec.~\ref{sec:Derivations_of_formulae_orthogonal}, we
consider the case where the pre- and post-selected states is
orthogonal.


\subsection{Non-orthogonal case}
\label{sec:Derivations_of_formulae_non_orthogonal}


When the norm $\left|\langle\psi_{f}|\psi_{i}\rangle\right|$ is
non-vanishing, the density matrix after the post-selection is
given by Eq.~(\ref{eq:detector-density-matrix-after-PS}).
Only through the property ${\bf A}^{2}=1$, the normalization
factor ${\cal Z}$ and the density matrix $\rho_{d}'$, which are
given by Eqs.~(\ref{eq:total-trace-of-PS-density-matrix-Z}) and 
(\ref{eq:detector-density-matrix-after-PS}), are reduced to the
following series
\begin{widetext}
\begin{eqnarray}
  {\cal Z}
  &=&
  1
  +
  \frac{1}{2}
  \sum_{n=1}^{+\infty} \frac{(-i2g)^{2n}}{(2n)!}\langle p^{2n}\rangle
  \left(
    1
    -
    \left|\langle{\bf A}\rangle_{w}\right|^{2}
  \right)
  +
  i \sum_{n=0}^{+\infty} \frac{(-i2g)^{2n+1}}{(2n+1)!}\langle p^{2n+1}\rangle
  \Im\langle{\bf A}\rangle_{w}
  ,
  \label{eq:kouchan-A2is1-9}
  \\
  \rho_{d}'
  &=&
  {\cal Z}^{-1}\left[
    \rho_{d}
    +
    \sum_{n=1}^{+\infty} \frac{(-ig)^{2n}}{(2n)!} 
    \left(
      \sum_{k=0}^{n} {}_{2n}C_{2k} p^{2n-2k}\rho_{d}p^{2k}
      -
      \left|\langle{\bf A}\rangle_{w}\right|^{2}
      \sum_{k=0}^{n-1} {}_{2n}C_{2k+1} p^{2n-2k-1}\rho_{d}p^{2k+1}
    \right)
  \right.
  \nonumber\\
  && \quad\quad\quad
  \left.
    +
    \sum_{n=0}^{+\infty} \frac{(-ig)^{2n+1}}{(2n+1)!} 
    \left(
      \langle{\bf A}\rangle_{w} 
      \sum_{k=0}^{n} {}_{2n+1}C_{2k} p^{2n+1-2k}\rho_{d}p^{2k}
      -
      \langle{\bf A}\rangle_{w}^{*} 
      \sum_{k=0}^{n} {}_{2n+1}C_{2k+1} p^{2n-2k}\rho_{d}p^{2k+1}
    \right)
  \right]
  .
  \label{eq:kouchan-A2is1-28-2}
\end{eqnarray}
From this density matrix (\ref{eq:kouchan-A2is1-28-2}), we can
evaluate the expectation values of $p$, $q$, $p^{2}$, and
$q^{2}$ after the post-selection as follows
\begin{eqnarray}
  {\cal Z} \langle p\rangle'
  &=&
  \langle p\rangle
  +
  \frac{1}{2}
  \left( 1 - \left|\langle{\bf A}\rangle_{w}\right|^{2} \right)
  \sum_{n=1}^{+\infty} \frac{(-i2g)^{2n}}{(2n)!} \langle p^{2n+1}\rangle
  +
  i \Im\langle{\bf A}\rangle_{w}
  \sum_{n=0}^{+\infty} \frac{(-i2g)^{2n+1}}{(2n+1)!} \langle p^{2n+2}\rangle
  \label{eq:kouchan-A2is1-30-2}
  ,
  \\
  {\cal Z}\langle q\rangle'
  &=&
  \langle q\rangle
  + g
  \left(
      \Re\langle{\bf A}\rangle_{w}
    + \Im\langle{\bf A}\rangle_{w} \langle(qp+pq)\rangle
  \right)
  + \frac{1}{2} i \Im\langle{\bf A}\rangle_{w}
  \sum_{n=1}^{+\infty} \frac{(-i2g)^{2n+1}}{(2n+1)!}
  \langle qp^{2n+1}+p^{2n+1}q\rangle
  \nonumber\\
  &&
  + \frac{1}{4} \left(
    1 - \left|\langle{\bf A}\rangle_{w}\right|^{2}
  \right)
  \sum_{n=1}^{+\infty} \frac{(-4g^{2})^{n}}{(2n)!}
  \langle qp^{2n} + p^{2n}q\rangle
  ,
  \label{eq:kouchan-A2is1-37}
  \\
  {\cal Z}\langle p^{2}\rangle'
  &=&
  \langle p^{2}\rangle
  +
  \frac{1}{2}
  \left(
    1 - \left|\langle{\bf A}\rangle_{w}\right|^{2}
  \right)
  \left(
    \sum_{n=1}^{+\infty} \frac{(-i2g)^{2n}}{(2n)!} \langle p^{2n+2}\rangle 
  \right)
  + i \Im\langle{\bf A}\rangle_{w} 
  \left(
    \sum_{n=0}^{+\infty} \frac{(-i2g)^{2n+1}}{(2n+1)!} \langle p^{2n+3}\rangle
  \right)
  \label{eq:kouchan-A2is1-54-2}
  ,
  \\
  {\cal Z}\langle q^{2}\rangle'
  &=&
  \langle q^{2}\rangle
  + g
  \left(
      2 \Re\langle{\bf A}\rangle_{w} \langle q\rangle
    +   \Im\langle{\bf A}\rangle_{w} \langle q^{2}p+pq^{2}\rangle
  \right)
  +
  \frac{g^{2}}{2} 
  \left(
      2 \left|\langle{\bf A}\rangle_{w}\right|^{2}
    -   \left(
      1 - \left|\langle{\bf A}\rangle_{w}\right|^{2}
    \right) \langle q^{2}p^{2}+p^{2}q^{2}\rangle
  \right)
  \nonumber\\
  &&
  +
  \frac{1}{2^{2}} 
  \left( 1 - \left|\langle{\bf A}\rangle_{w}\right|^{2} \right)
  \sum_{n=2}^{+\infty} \frac{(-i2g)^{2n}}{(2n)!}
  \left(
    \langle q^{2}p^{2n}+p^{2n}q^{2}\rangle
    + \frac{1}{2} (2n)(2n-1) \langle p^{2n-2}\rangle
  \right)
  \nonumber\\
  &&
  +
  \frac{i}{2} \Im\langle{\bf A}\rangle_{w} 
  \sum_{n=1}^{+\infty} \frac{(-2ig)^{2n+1}}{(2n+1)!}
  \left(
      \langle q^{2}p^{2n+1}+p^{2n+1}q^{2}\rangle
    + \frac{1}{2} (2n+1)(2n) \langle p^{2n-1}\rangle
  \right)
  \label{eq:kouchan-A2is1-81}
\end{eqnarray}
where $\langle *\rangle':=\TR_{d}\left(*\rho_{d}'\right)$.
Further, the probability densities
$\langle p|\rho_{d}'|p\rangle$ in $p$-space and 
$\langle q|\rho_{d}'|q\rangle$ in $q$-space are given by 
\begin{eqnarray}
  {\cal Z}\langle p|\rho_{d}'|p\rangle
  &=&
  \left[
    1
    + \frac{1}{2}
    \left(
      1 - \left|\langle{\bf A}\rangle_{w}\right|^{2}
    \right)
    \left(
      \cos(2gp) - 1
    \right)
    + \Im\langle{\bf A}\rangle_{w} \sin(2gp)
  \right]
  \langle p|\rho_{d}|p\rangle 
  ,
  \label{eq:kouchan-A2is1-99}
  \\
  {\cal Z}\langle q|\rho_{d}'|q\rangle
  &=&
  \left|\langle q|\phi\rangle\right|^{2}
  +
  {\cal I}
  +
  {\cal J}
  \label{eq:kouchan-A2is1-104}
  ,
\end{eqnarray}
where ${\cal I}$ and ${\cal J}$ are defined by 
\begin{eqnarray}
  {\cal I}
  &:=&
  \sum_{n=1}^{+\infty} \frac{g^{2n}}{(2n)!} 
  \left\{
    \sum_{k=0}^{n} {}_{2n}C_{2k}
    \left(
      \frac{\partial^{2n-2k}}{\partial q^{2n-2k}}\langle q|\phi\rangle
    \right)
    \left(
      \frac{\partial^{2k}}{\partial q^{2k}}\langle q|\phi\rangle
    \right)^{*}
  \right.
  \nonumber\\
  && \quad\quad\quad\quad\quad\quad\quad\quad\quad\quad
  \left.
    +
    \left|\langle{\bf A}\rangle_{w}\right|^{2}
    \sum_{k=0}^{n-1} {}_{2n}C_{2k+1} 
    \left(
      \frac{\partial^{2n-2k-1}}{\partial q^{2n-2k-1}}\langle q|\phi\rangle
    \right)
    \left(
      \frac{\partial^{2k+1}}{\partial q^{2k+1}}\langle q|\phi\rangle
    \right)^{*}
  \right\}
  \label{eq:calI-def}
  , \\
  {\cal J}
  &:=&
  -
  \sum_{n=0}^{+\infty} \frac{g^{2n+1}}{(2n+1)!} 
  \left\{
    \langle{\bf A}\rangle_{w} 
    \sum_{k=0}^{n} {}_{2n+1}C_{2k}
    \left(
      \frac{\partial^{2n+1-2k}}{\partial q^{2n+1-2k}}\langle q|\phi\rangle
    \right) \left(
      \frac{\partial^{2k}}{\partial q^{2k}}\langle q|\phi\rangle
    \right)^{*}
  \right.
  \nonumber\\
  && \quad\quad\quad\quad\quad\quad\quad\quad
  \left.
    +
    \langle{\bf A}\rangle_{w}^{*} 
    \sum_{k=0}^{n} {}_{2n+1}C_{2k+1}
    \left(
      \frac{\partial^{2n-2k}}{\partial q^{2n-2k}}\langle q|\phi\rangle
    \right)
    \left(
      \frac{\partial^{2k+1}}{\partial q^{2k+1}}\langle q|\phi\rangle
    \right)^{*}
  \right\}
  .
  \label{eq:calJ-def}
\end{eqnarray}
\end{widetext}


When the initial state of the detector is zero-mean value
Gaussian (\ref{eq:instate-of-detector-is-Gaussian}), the moments
of $p$ are given by Eqs.~(\ref{eq:instate-p-moments-in-Gaussian}).
In this case, the normalization ${\cal Z}$
[Eq.~(\ref{eq:kouchan-A2is1-9})] is given by
Eq.~(\ref{eq:kouchan-A2is1-10-2}).  
Similar calculations with the properties 
\begin{eqnarray}
  \label{eq:kouchan-A2is1-orthogonal-2.4.5}
  \langle(qp^{n}+p^{n}q)\rangle 
  &=&
  0
  \quad \mbox{for} \quad n\geq 1
  ,
  \\
  \langle q^{2}p^{2n}+p^{2n}q^{2}\rangle
  &=&
  - \frac{2n-1}{2} \frac{(2n-1)!!}{(2a)^{n-1}}
  \label{eq:kouchan-A2is1-90-0}
  , \\
  \langle q^{2}p+pq^{2}\rangle
  &=&
  \langle q^{2}p^{2n+1}+p^{2n+1}q^{2}\rangle
  \nonumber\\
  &=& 0
  , \quad \mbox{for} \quad n\geq 1
  \label{eq:kouchan-A2is1-90}
\end{eqnarray}
of the zero mean-value Gaussian state lead to the expectation 
values of $p$ and $q$  [Eq.~(\ref{eq:kouchan-A2is1-52}) and
(\ref{eq:kouchan-A2is1-53})] after the post-selection and the
variances in $p$ and $q$ [Eq.~(\ref{eq:kouchan-A2is1-94}) and
(\ref{eq:kouchan-A2is1-95})] after the post-selection.
We also note that the derivation of the probability density
(\ref{eq:kouchan-A2is1-101}) in $p$-space is straight forward,
while the derivation of Eq.~(\ref{eq:kouchan-A2is1-141}) is
non-trivial. 
Therefore, we only explain the derivation
Eq.~(\ref{eq:kouchan-A2is1-141}).


The initial state $\langle q|\phi\rangle$ of the detector is
derived from the Fourier transformation of
Eq.~(\ref{eq:instate-of-detector-is-Gaussian}):
\begin{eqnarray}
  \label{eq:instate-Gaussian-in-q-space}
  \langle q|\phi\rangle
  &=&
  \left(\frac{s}{\pi g^{2}}\right)^{1/4}
  \exp\left[
    - \frac{s}{2}\left(\frac{q}{g}\right)^{2}
  \right]
  .
\end{eqnarray}
From the definition of the Hermite
polynomial~\cite{I.S.Gradshteyn-I.M.Ryzhik-2000}: 
\begin{eqnarray}
  \label{eq:kouchan-A2is1-107}
  H_{n}(x)
  :=
  (-1)^{n} e^{x^{2}/2}\frac{d^{n}}{dx^{n}}(e^{-x^{2}/2})
  ,
\end{eqnarray}
we easily obtain 
\begin{eqnarray}
  \frac{\partial^{n}}{\partial q^{n}}\langle q|\phi\rangle
  =
  (-g)^{-n} s^{n/2}
  H_{n}\left(
    \frac{\sqrt{s}q}{g}
  \right)
  \langle q|\phi\rangle
  .
  \label{eq:kouchan-A2is1-112}
\end{eqnarray}
This formula (\ref{eq:kouchan-A2is1-112}) is used to evaluate
the derivative of the initial wave function
(\ref{eq:instate-Gaussian-in-q-space}) in
Eqs.~(\ref{eq:calI-def}) and (\ref{eq:calJ-def}).


To evaluate $\langle q|\rho_{d}'|q\rangle$ through
Eq.~(\ref{eq:kouchan-A2is1-104}), we first consider the second
term ${\cal I}$ in Eq.~(\ref{eq:kouchan-A2is1-104}): 
\begin{widetext}
\begin{eqnarray}
  {\cal I}
  &=&
  \langle q|\rho_{d}|q\rangle
  \sum_{n=1}^{+\infty} \frac{s^{n}}{(2n)!}
  \left\{
    \sum_{k=0}^{n} {}_{2n}C_{2k}
    H_{2n-2k}\left(\frac{\sqrt{s}q}{g}\right)
    H_{2k}\left(\frac{\sqrt{s}q}{g}\right)
  \right.
  \nonumber\\
  && \quad\quad\quad\quad\quad\quad\quad\quad\quad\quad
  \left.
    +
    \left|\langle{\bf A}\rangle_{w}\right|^{2}
    \sum_{k=0}^{n-1} {}_{2n}C_{2k+1} 
    H_{2n-2k-1}\left(\frac{\sqrt{s}q}{g}\right)
    H_{2k+1}\left(\frac{\sqrt{s}q}{g}\right)
  \right\}
  \label{eq:A12}
\end{eqnarray}
\end{widetext}


Here, we note that the Hermite polynomial
(\ref{eq:kouchan-A2is1-107}) is an even function of $x$ if the 
index $n$ is even and an odd function of $x$ if the index $n$ is 
odd~\cite{I.S.Gradshteyn-I.M.Ryzhik-2000}, i.e.,  
\begin{eqnarray}
  &&
  \label{eq:kouchan-A2is1-114-even}
  H_{2n}(-x) = H_{2n}(x)
  ,\\
  &&
  \label{eq:kouchan-A2is1-114-odd}
  H_{2n+1}(-x)=-H_{2n+1}(x),
  \\
  \label{eq:kouchan-A2is1-zero-point}
  &&
  H_{2k}(0)=(-1)^{k}(2k-1)!!, \;\; H_{2k+1}(0)=0.
\end{eqnarray}
Further, we also note the sum rule of the Hermite
polynomial~\cite{I.S.Gradshteyn-I.M.Ryzhik-2000}: 
\begin{eqnarray}
  H_{n}(x+y)
  =
  \frac{1}{2^{n/2}} \sum_{r=0}^{n}{}_{n}C_{r}H_{n-r}(\sqrt{2}x)H_{r}(\sqrt{2}y) 
  .
  \nonumber\\
  \label{eq:kouchan-A2is1-115}
\end{eqnarray}
Through these formulae
(\ref{eq:kouchan-A2is1-114-even})--(\ref{eq:kouchan-A2is1-115}),
we easily obtain
\begin{eqnarray}
  &&
  \sum_{r=0}^{k}{}_{2k}C_{2r}H_{2k-2r}(\sqrt{2}x)H_{2r}(\sqrt{2}x) 
  \nonumber\\
  &=&
  2^{k-1}
  \left(
    H_{2k}(2x)
    +
    (-1)^{k}(2k-1)!!
  \right)
  \label{eq:kouchan-A2is1-128}
  , \\
  &&
  \sum_{r=0}^{k-1}{}_{2k}C_{2r+1}H_{2k-2r-1}(\sqrt{2}x)H_{2r+1}(\sqrt{2}x)
  \nonumber\\
  &=&
  2^{k-1}\left(H_{2k}(2x) - (-1)^{k}(2k-1)!!\right)
  \label{eq:kouchan-A2is1-129}
  , \\
  &&
  \sum_{r=0}^{k}{}_{2k+1}C_{2r}H_{2k+1-2r}(\sqrt{2}x)H_{2r}(\sqrt{2}x)
  \nonumber\\
  &=&
  2^{k-1/2}
  H_{2k+1}(2x)
  \label{eq:kouchan-A2is1-130}
  , \\
  &&
  \sum_{r=0}^{k}{}_{2k+1}C_{2r+1}H_{2k-2r}(\sqrt{2}x)H_{2r+1}(\sqrt{2}x)
  \nonumber\\
  &=&
  2^{k-1/2}
  H_{2k+1}(2x)
  \label{eq:kouchan-A2is1-131}
  .
\end{eqnarray}
Through the formulae
(\ref{eq:kouchan-A2is1-128})--(\ref{eq:kouchan-A2is1-131}), 
Eq.~(\ref{eq:A12}) is given by 
\begin{eqnarray}
  {\cal I}
  &=&
  \frac{1}{2}\left|\langle q|\phi\rangle\right|^{2}
  \left[
    \sum_{n=0}^{+\infty}
    \frac{(2s)^{n}}{(2n)!}
    H_{2n}\left(\frac{\sqrt{2s}q}{g}\right)
  \right.
  \nonumber\\
  && \quad\quad\quad\quad\quad
  \left.
    + \left(
      1 - \left|\langle{\bf A}\rangle_{w}\right|^{2}
    \right)e^{-s}
    - 2
  \right.
  \nonumber\\
  && \quad\quad\quad\quad\quad
  \left.
    + \left|\langle{\bf A}\rangle_{w}\right|^{2}
    \left(
      \sum_{n=0}^{+\infty}
      \frac{(2s)^{n}}{(2n)!} H_{2n}\left(\frac{\sqrt{2s}q}{g}\right)
    \right)
  \right]
  .
  \nonumber\\
  \label{eq:kouchan-A2is1-132}
\end{eqnarray}


Here, we note the formulae~\cite{I.S.Gradshteyn-I.M.Ryzhik-2000}: 
\begin{eqnarray}
  \sinh(tx)
  &=&
  e^{t^{2}/2} \sum_{n=0}^{\infty}H_{2n+1}(x)\frac{t^{2n+1}}{(2n+1)!}
  \label{eq:kouchan-A2is1-134}
  , \\
  \cosh(tx)
  &=&
  e^{t^{2}/2} \sum_{n=0}^{\infty}H_{2n}(x)\frac{t^{2n}}{(2n)!}
  \label{eq:kouchan-A2is1-136}
  .
\end{eqnarray}
Through the formula (\ref{eq:kouchan-A2is1-136}),
\begin{eqnarray}
  {\cal I}
  &=&
  \left|\langle q|\phi\rangle\right|^{2} e^{-s}
  \left[
    \cosh^{2}\left(\frac{sq}{g}\right)
    - e^{s}
  \right.
  \nonumber\\
  && \quad\quad\quad\quad\quad\quad
  \left.
    + \left|\langle{\bf A}\rangle_{w}\right|^{2} \sinh^{2}\left(\frac{sq}{g}\right)
  \right]
  .
  \label{eq:kouchan-A2is1-135-2}
\end{eqnarray}
Using Eq.~(\ref{eq:kouchan-A2is1-134}), the similar evaluation
of the final term ${\cal J}$ in Eq.~(\ref{eq:kouchan-A2is1-104})
yields
\begin{eqnarray}
  {\cal J}
  &=&
  \left|\langle q|\phi\rangle\right|^{2}
  \Re\langle{\bf A}\rangle_{w} 
  e^{-s} \sinh\left(\frac{2sq}{g}\right)
  \label{eq:kouchan-A2is1-137}
  .
\end{eqnarray}


Through Eqs.~(\ref{eq:kouchan-A2is1-135-2}) and
(\ref{eq:kouchan-A2is1-137}), we can evaluate
Eq.~(\ref{eq:kouchan-A2is1-104}) and the probability density in
$q$-space is given by Eq.~(\ref{eq:kouchan-A2is1-141}) with the
initial probability density in $q$-space
(\ref{eq:initial-Gaussian-prob-density-q-space}).


\subsection{Orthogonal case}
\label{sec:Derivations_of_formulae_orthogonal}


In the orthogonal weak measurements 
$\langle\psi_{f}|\psi_{i}\rangle=0$, the orthogonal weak value
is trivial as shown in
Eq.~(\ref{eq:kouchan-A2is1-orthogonal-14}). 
Through the orthogonal weak value
(\ref{eq:kouchan-A2is1-orthogonal-14}), the normalization
constant ${\cal Z}_{o}$ defined by
Eq.~(\ref{eq:S.Wu-Y.Li-2010-22}) is given by  
\begin{eqnarray}
  \label{eq:kouchan-A2is1-orthogonal-16}
  {\cal Z}_{o}
  &=&
  1
  +
  2
  \sum_{n=1}^{+\infty} \frac{(-4g^{2})^{n}}{(2n+2)!}
  \frac{\langle p^{2n+2}\rangle}{\langle p^{2}\rangle}
  .
\end{eqnarray}
The density matrix of the detector after the post-selection is
given by Eq.~(\ref{eq:kouchan-A2is1-orthogonal-2.3.2}).


The expectation value of $p$, $q$, $p^{2}$, and $q^{2}$ after
the post-selection are evaluated as 
\begin{eqnarray}
  {\cal Z}_{o} \langle p^{2}\rangle \langle p\rangle'
  &=&
  {\cal Z}_{o} \langle p^{2}\rangle \TR\left(p\rho_{d}'\right)
  \nonumber\\
  &=&
  \langle p^{3}\rangle
  + 2
  \sum_{n=1}^{+\infty} \frac{(-i2g)^{2n}}{(2n+2)!}
  \langle p^{2n+3}\rangle
  ,
  \label{eq:kouchan-A2is1-orthogonal-2.4.1}
  \\
  {\cal Z}_{o} \langle p^{2}\rangle \langle q\rangle'
  &=&
  {\cal Z}_{o} \langle p^{2}\rangle \TR\left(q\rho_{d}'\right)
  \nonumber\\
  &=&
  \frac{1}{2} \langle qp^{2}+p^{2}q\rangle
  \nonumber\\
  &&
  + \sum_{n=1}^{+\infty} \frac{(-i2g)^{2n}}{(2n+2)!}
  \nonumber\\
  && \quad\quad
  \times
  \langle qp^{2n+2}+p^{2n+2}q\rangle
  ,
  \label{eq:kouchan-A2is1-orthogonal-2.4.3}
  \\
  {\cal Z}_{o} \langle p^{2}\rangle \langle p^{2}\rangle'
  &=&
  {\cal Z}_{o} \langle p^{2}\rangle \TR\left(p^{2}\rho_{d}'\right)
  \nonumber\\
  &=&
  \langle p^{4}\rangle
  + 2 \sum_{n=1}^{+\infty} \frac{(-4g^{2})^{n}}{(2n+2)!} \langle p^{2n+4}\rangle
  ,
  \label{eq:kouchan-A2is1-orthogonal-2.4.9}
  \\
  {\cal Z}_{o} \langle p^{2}\rangle \langle q^{2}\rangle'
  &=&
  {\cal Z}_{o} \langle p^{2}\rangle \TR\left[q^{2}\rho_{d}'\right]
  \nonumber\\
  &=&
  1
  + \frac{1}{2} \langle q^{2}p^{2}+p^{2}q^{2}\rangle
  \nonumber\\
  &&
  + \frac{1}{2} \sum_{n=1}^{+\infty} \frac{(-4g^{2})^{n}}{(2n+2)!} \left(
    2 \langle q^{2}p^{2n+2}+p^{2n+2}q^{2}\rangle
  \right.
  \nonumber\\
  && \quad\quad\quad
  \left.
    + (2n+2) (2n+1) \langle p^{2n}\rangle
  \right)
  .
  \label{eq:kouchan-A2is1-orthogonal-2.4.13}
\end{eqnarray}
From the density matrix
(\ref{eq:kouchan-A2is1-orthogonal-2.3.2}), we can directly
obtain the probability density in $p$-space as
\begin{eqnarray}
  \langle p|\rho_{d}'|p\rangle
  =
  \frac{1}{{\cal Z}_{o} \langle p^{2}\rangle}
  \frac{1}{4g^{2}}
  \left(1 - \cos(2gp) \right)
  \langle p|\rho_{d}|p\rangle
  .
  \label{eq:kouchan-A2is1-orthogonal-2.6.2}
\end{eqnarray}
\begin{widetext}
On the other hand, we also obtain the probability density in
$q$-space as 
\begin{eqnarray}
  {\cal Z}_{o} \langle p^{2}\rangle \langle q|\rho_{d}'|q\rangle
  &=&
  \langle q|p\rho_{d}p|q\rangle
  +
  \sum_{n=1}^{+\infty} \frac{(-ig)^{2n}}{(2n+2)!}
  \sum_{k=0}^{n} {}_{2n+2}C_{2k+1}
  \langle q|p^{2(n-k)+1}\rho_{d}p^{2k+1}|q\rangle
  .
  \label{eq:kouchan-A2is1-orthogonal-2.7.1}
\end{eqnarray}
since we choose the initial state of the detector as a pure state 
$\rho_{d}=|\phi\rangle\langle\phi|$,
Eq.~(\ref{eq:kouchan-A2is1-orthogonal-2.7.1}) yields
\begin{eqnarray}
  \langle q|\rho_{d}'|q\rangle
  &=&
  \frac{1}{{\cal Z}_{o} \langle p^{2}\rangle}
  \left[
    \left(
      \frac{\partial}{\partial q}\langle q|\phi\rangle
    \right)
    \left(\frac{\partial}{\partial q}\langle q|\phi\rangle\right)^{*}
  \right.
  \nonumber\\
  && \quad\quad\quad\quad
  \left.
    +
    \sum_{n=1}^{+\infty} \frac{g^{2n}}{(2n+2)!}
    \sum_{k=0}^{n} {}_{2n+2}C_{2k+1} 
    \left(
      \frac{\partial^{2(n-k)+1}}{\partial q^{2(n-k)+1}}\langle q|\phi\rangle
    \right)
    \left(
      \frac{\partial^{2k+1}}{\partial q^{2k+1}}\langle q|\phi\rangle
    \right)^{*}
  \right]
  .
  \label{eq:kouchan-A2is1-orthogonal-2.7.3}
\end{eqnarray}
\end{widetext}


When the initial state of the detector is Gaussian
(\ref{eq:instate-of-detector-is-Gaussian}), we use 
Eqs.~(\ref{eq:instate-p-moments-in-Gaussian}) and
(\ref{eq:kouchan-A2is1-orthogonal-2.4.5})--(\ref{eq:kouchan-A2is1-90}).
Then, the expectation values of $p$ and $q$ after the
post-selection are trivial as shown in 
Eqs.~(\ref{eq:kouchan-A2is1-orthogonal-2.4.2}) and
(\ref{eq:kouchan-A2is1-orthogonal-2.4.6}). 
Since the expectation values of $p$ and $q$ after the
post-selection vanish, the expectation values of $p^{2}$ and
$q^{2}$ themselves represent the variances in $p$ and $q$ after
the post-selection.
Then we obtain Eqs.~(\ref{eq:kouchan-A2is1-orthogonal-2.4.11})
and (\ref{eq:kouchan-A2is1-orthogonal-2.4.15}).
Furthermore, the probability density
(\ref{eq:kouchan-A2is1-orthogonal-2.6.2}) in $p$-space trivially
yields Eq.~(\ref{eq:kouchan-A2is1-orthogonal-2.6.4}).
However, the expression of the probability density
(\ref{eq:kouchan-A2is1-orthogonal-2.7.6}) in $q$-space requires
the non-trivial derivation from
Eq.~(\ref{eq:kouchan-A2is1-orthogonal-2.7.3}). 
Therefore, we briefly explain the derivation of
Eq.~(\ref{eq:kouchan-A2is1-orthogonal-2.7.6}) below.


Since our initial state of the detector is a zero mean-value
Gaussian (\ref{eq:instate-Gaussian-in-q-space}), we also apply
the formula (\ref{eq:kouchan-A2is1-112}).
Substituting Eq.~(\ref{eq:kouchan-A2is1-112}) into
Eq.~(\ref{eq:kouchan-A2is1-orthogonal-2.7.3}), we obtain 
\begin{widetext}
\begin{eqnarray}
  {\cal Z}_{o} \langle p^{2}\rangle \langle q|\rho_{d}'|q\rangle
  &=&
  \left|\langle q|\phi\rangle\right|^{2}
  \frac{s^{2}q^{2}}{g^{4}}
  \nonumber\\
  &&
  +
  \left|\langle q|\phi\rangle\right|^{2}
  \frac{s}{g^{2}}
  \sum_{n=1}^{+\infty} \frac{s^{n}}{(2n+2)!}
  \sum_{k=0}^{n} {}_{2n+2}C_{2k+1} 
  H_{2n-2k+1}\left(\frac{\sqrt{s}q}{g}\right)
  H_{2k+1}\left(\frac{\sqrt{s}q}{g}\right)
  .
\end{eqnarray}
\end{widetext}
Through formulae (\ref{eq:kouchan-A2is1-136}) and
(\ref{eq:kouchan-A2is1-129}), the calculations similar to the 
derivation (\ref{eq:kouchan-A2is1-141}) yields
(\ref{eq:kouchan-A2is1-orthogonal-2.7.6}).



\end{document}